\documentclass[pra,aps,twocolumn,groupedaddress,nofootinbib,noshowpacs,preprintnumbers,longbibliography,floatfix]{revtex4-2}
\usepackage[utf8]{inputenc}
\usepackage{graphicx}
\usepackage{float}
\usepackage{amssymb}
\usepackage{amsmath}  
\usepackage{mathtools}
\usepackage{dsfont}
\usepackage{array}
\usepackage{bm,fixmath}
\usepackage{subfigure}
\usepackage{mathrsfs}
\usepackage{pifont}
\usepackage{multirow}
\usepackage{upgreek}
\usepackage{xcolor}
\usepackage{bm}
\usepackage{bbm}
\usepackage{qcircuit}
\usepackage{physics}
\usepackage{comment}
\usepackage{appendix}
\usepackage{verbatim}
\usepackage{slashed}
\usepackage[pdftex,
            colorlinks,
            linkcolor=red,
            citecolor=blue,
            urlcolor=black]{hyperref}
\usepackage{cleveref}
\usepackage{tikz}
\usetikzlibrary{zx-calculus}

\newsavebox{\mstrut}
\newcommand{\dbra}[1]{%
    \sbox{\mstrut}{\(#1\)}%
    \mathinner{\left\langle\kern-0.25\ht\mstrut\left\langle{#1}\right|\mkern-2mu\right.}%
}
\newcommand{\dket}[1]{%
    \sbox{\mstrut}{\(#1\)}%
    \mathinner{\left.\mkern-2mu\left|{#1}\right\rangle\kern-0.25\ht\mstrut\right\rangle}%
}

\usetikzlibrary{calc}
\makeatletter
\DeclareRobustCommand\rvdots{%
\vbox{\baselineskip4\p@\lineskiplimit\z@\kern-\p@\hbox{.}\hbox{.}\hbox{.}}}

\DeclareRobustCommand\elevdots{%
\vbox{\baselineskip4\p@\lineskiplimit\z@\kern-\p@\hbox{.}\hbox{.}\hbox{.}\vspace{-0.1cm}}}

\begin{document}

\title{Flexible entangled state generation in linear optics}

\author{Brendan Pankovich}
\author{Alex Neville}
\author{Angus Kan}
\author{Srikrishna Omkar}
\author{Kwok Ho Wan}
\author{Kamil Brádler}
\email{kamil@orcacomputing.com}
\affiliation{ORCA Computing}

\begin{abstract}
Fault-tolerant quantum computation can be achieved by creating constant-sized, entangled resource states and performing entangling measurements on subsets of their qubits.
Linear optical quantum computers can be designed based on this approach, even though entangling operations at the qubit level are non-deterministic in this platform.
Probabilistic generation and measurement of entangled states must be pushed beyond the required threshold by some combination of scheme optimisation, introduction of redundancy and auxiliary state assistance.
We report progress in each of these areas.
We explore multi-qubit fusion measurements on dual-rail photonic qubits and their role in measurement-based resource state generation, showing that it is possible to boost the success probability of photonic GHZ state analysers with single photon auxiliary states.
By incorporating generators of basic entangled ``seed" states, we provide a method that simplifies the process of designing and optimising generators of complex, encoded resource states by establishing links to ZX diagrams.

\end{abstract}

\maketitle

\section{Introduction}

It has long been known that universal quantum computation can be achieved with a source of single photons, linear optics and photon number detection~\cite{Knill2001ASF}.
Approaches best suited to the flying qubits present in linear optical quantum computers are measurement based~\cite{briegel2001persistent, raussendorf2001one, raussendorf2003mbqc}, and rely on generating entangled states as a resource.

Entangled resource states allow for modular construction of fault-tolerant measurement-based architectures~\cite{raussendorf2007topological,brown2020universal}, for instance by first combining small resource states through entangling measurements into a cluster state, and then performing single-qubit measurements~\cite{li2015resource,gimeno2015from,omkar2022all}. 
More streamlined approaches achieve fault-tolerance by directly performing small entangling measurements on a collection of constant-sized resource states \cite{bartolucci2021fusion,encodedghz,sahay2022tailoring}, forgoing the generation of any particularly large entangled state.
Beyond quantum computation, resource states have found applications in quantum communication, where they enable fault-tolerant long-distance entanglement distribution~\cite{ewert2016ultrafast,ewert2017ultrafast,lee2019fundamental}, and in quantum metrology~\cite{friis2017flexible,kwon2019Nonclassicality,shettell2020graph}.

In addition to acting on prepared resource states, entangling measurements are central to generating the resource states themselves.
Methods for generating the smallest, most basic of resource states, known as ``seed states"~\cite{bartolucci2021creation}, from single photons rely on entangling measurements to operate in a useful way~\cite{zhang2007demonstration, varnava2008howgood}.
Furthermore, generating larger resource states from seed states typically depends on two-qubit entangling measurements, such as Bell state measurements~\cite{browne2005resource-efficient,gimeno2015from}.

Useful entanglement generating operations are intrinsically non-deterministic using linear optics, and their success probabilities are degraded by photon loss.
While these facts do not rule out useful linear optical quantum computing, they do necessitate particularly careful design of state generation and measurement schemes.
Significant effort has been made to increase their efficiencies via auxiliary assistance~\cite{grice2011arbitrarily, ewert2014efficient, bartolucci2021creation}, adding redundancy with error-correcting codes~\cite{ewert2016ultrafast,lee2019fundamental,schmidt2019efficiencies,bartolucci2021fusion,hilaire2023linear,bell2023optimizing}, and general device optimisation~\cite{bartolucci2021creation, fldzhyan2021compact}.

Designing schemes for generating larger resource states, such as those with encoded qubits, can be challenging. Existing techniques~\cite{lee2023graphtheoretical, ewert2017ultrafast, li2015resource} typically involve finding an LU-equivalent graph state for a target stabilizer state and breaking that graph state into smaller ones that are connected via fusion measurements.
However, these techniques apply to specific classes of graph states and are limited in efficiency due to their reliance on single-qubit and fusion measurements.
The need for something beyond the graph state-fusion paradigm is well illustrated in the case of inner-encoded graph states. While the description of the logical state and inner code may be quite simple, representing this type of concatenated structure as a graph state on the physical qubits reveals a complex web of correlations between qubits, which can lead to convoluted fusion-based generation schemes.

In this paper, we introduce a framework that simplifies reasoning with linear-optical resource state generation by using connections between specific measurement devices and ZX diagrams~\cite{coeke2008interacting, coecke2011interacting}.
By shifting the focus from the usual two-qubit entangling measurements, we also propose auxiliary assistance schemes for general GHZ analyser devices which provide a relative performance boost per auxiliary photon which grows with the size of the measurement.

\section{Linear-optical state-generation tools}

We consider spatial dual-rail (DR) encoded photonic qubits, for which there is the following correspondence between qubits and two-mode Fock states:
\begin{equation}
\begin{split}
    \ket{0} &= \dket{10}\\
    \ket{1} &= \dket{01},
\end{split}
\label{eq:qubit-fock}
\end{equation}
where we use $\ket{\cdot}$ to refer to a qubit state and $\dket{\cdot}$ to a multimode Fock state.

The purpose of this section is to outline a set of basic tools for devising schemes for generating complicated entangled states with linear optics, seed states and single photon detectors.
These tools are \emph{type-I $n$-fusion} which generalises two-qubit type-I fusion, and \emph{entangled state analysers}, which can be used to implement type-II $n$-fusion.

We will make associations between each of these tools with certain ZX diagrams, which will enable us to make use of ZX calculus for the intuitive redesign and optimisation of state generation devices.

When graphically representing linear optical devices, we follow~\cite{bartolucci2021creation} in representing beamsplitters in the way that CZ gates typically are for quantum circuit diagrams.
More concretely, we use
\begin{equation*}
\vcenter{\hbox{\resizebox{0.05\textwidth}{!}{\includegraphics{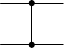}}}}
\end{equation*}
to represent a 50:50 beamsplitter applied to two optical modes, which correspond to the horizontal lines.

\subsection{Entangled-state analysers}

The state analyser devices that we study in this paper employ photon number resolving detector (PNRD) arrays coupled to linear optical networks (LONs), which perform destructive measurements. 
The action of a bare array of $m$ PNRDs can be described by a generalised measurement with Kraus operators of the form
\begin{equation}
    \bigotimes_{i=1}^{m}\dket{0_i}\hspace{-0.1cm}\dbra{r_i} = \dket{\bm{0}}\hspace{-0.1cm}\dbra{\bm{r}}
\end{equation}
where $\bm{r} = (r_1, \dots, r_m)$ is referred to as the detector (``click") pattern. In the $n$-photon subspace we have $\sum_{i}r_i = n$.

A LON with scattering matrix $U$ preceding the PNRD array transforms the Kraus operators such that
\begin{equation}
    \dket{\bm{0}}\hspace{-0.1cm}\dbra{\bm{r}}\mapsto \dket{\bm{0}}\hspace{-0.1cm}\dbra{\bm{r}} \mathcal{U}(U)
\end{equation}
where $\mathcal{U}(U)$ the multiphoton transformation associated with $U$, the elements of which are proportional to permanents of submatrices of $U$~\cite{scheel2004permanents}.
When the measurement device is operating on some subspace of input states---as is the case for idealised DR qubits---an additional projection onto that subspace is made.
As such, the Kraus operator associated with detector pattern $\bm{r}$ for an $n$-qubit measurement device restricted to $n$-qubit DR inputs can be written as:
\begin{equation}
  K_{\bm{r}} = \dket{\bm{0}}\hspace{-0.1cm}\dbra{\bm{r}} \mathcal{U} P_\mathrm{DR}^{\otimes n}.
  \label{eq:evolved-projected-kraus}
\end{equation}
where $P_\mathrm{DR} = \dket{10}\hspace{-0.1cm}\dbra{10} + \dket{01}\hspace{-0.1cm}\dbra{01}$, and the association between qubits and multimode Fock states can be made according to Eq.\ \eqref{eq:qubit-fock}.
As the measured photons always end up in the vacuum state of a mode which is subsequently discarded, we will use a shorthand notation in which Kraus operators take the form $\dbra{\cdot}$, or $\bra{\cdot}$ if a projection has been made onto the DR subspace.

When at least one $K_{\bm{r}}$ for a given device corresponds to a DR entangled state, we refer to that device as an entangled state analyser.

\subsubsection{Bell state analyser}
\label{subsubsec:bsms}

\begin{figure}[htp]
    \centering
    \includegraphics[width = 0.2\textwidth]{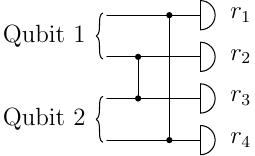}

    \caption{A Bell state analyser device composed of a linear optical network and single photon detectors. When acting on 2 dual rail input qubits, a measurement pattern $\bm{r}=(r_1, r_2, r_3, r_4)$ containing $r_i \in \{0,1\} \,\, \forall \,\,i$ indicates a destructive projection onto a specific dual rail Bell state. Other measurement patterns correspond to projections onto specific separable states.}
    \label{fig:BSM_circuit}
\end{figure}

An example of a Bell state analyser device~\cite{michler1996interferometric} is depicted in Fig.~\ref{fig:BSM_circuit}.
By applying Eq.\ \eqref{eq:evolved-projected-kraus} with a projection of the inputs onto the two-qubit DR subspace for each detector pattern, it is straightforward to show that this device performs a measurement described by the Kraus operators 
\begin{equation}
\begin{split}
    K_{(1,1,0,0)} = K_{(0,0,1,1)} &= \frac{1}{\sqrt{2}} \times \frac{1}{\sqrt{2}} \left( \bra{00} + \bra{11} \right) \\
    K_{(1,0,1,0)} = K_{(0,1,0,1)} &= \frac{1}{\sqrt{2}} \times \frac{1}{\sqrt{2}} \left( \bra{00} - \bra{11} \right) \\
    K_{(2,0,0,0)} = K_{(0,0,0,2)} &= \frac{1}{\sqrt{2}} \bra{01} \\
    K_{(0,2,0,0)} = K_{(0,0,2,0)}  &= \frac{1}{\sqrt{2}} \bra{10},
\end{split}
\label{eq:bsm-kraus}
\end{equation}
up to global phases.

Given a maximally mixed input, each of the eight possible detector patterns occur with equal probability, and four of them---$(1,1,0,0), (0,0,1,1), (1,0,1,0), (0,1,0,1)$---correspond to projection onto a Bell state.
The success probability of this device is therefore said to be $P_\mathrm{S} = 1/2$, which has been shown to be optimal for linear optical schemes without auxiliary assistance~\cite{calsamiglia2001maximum}.

By restricting ourselves to successful outcomes, it is possible to provide an alternative representation of the Bell state analyser device in terms of the ZX diagram
\begin{equation}\label{eq:bsm-zx}
\vcenter{\hbox{\resizebox{0.065\textwidth}{!}{\includegraphics{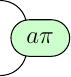}}}}
\end{equation}
where $a \in \{ 0,1 \}$ depends on the observed detector pattern, such that if $\bm{r} = (1, 1, 0, 0)$ or $\bm{r}= (0, 0, 1, 1)$ is observed, then $a=0$, and if $\bm{r} = (1, 0, 1, 0)$ or $\bm{r} = (0, 1, 0, 1)$ is observed, then $a=1$.

An important use case for linear optical Bell state measurements is in the generation of particular long-range quantum correlations which can allow for fault-tolerant quantum computation to be performed~\cite{raussendorf2007topological, bartolucci2021fusion}.
When considered in this context, successful Bell state analyser outcomes are often referred to as (type-II) fusion~\cite{browne2005resource-efficient} operations, because of their utility in merging two disconnected graph states into a single graph state.

As an instructive example~\cite{bartolucci2021creation}, consider applying the Bell state analyser device to the two isolated qubits of the state
\begin{equation*}
\begin{split}
\frac{1}{\sqrt{2}}\left( \ket{\psi_1}\ket{0} + \ket{\phi_1}\ket{1}\right) \otimes \frac{1}{\sqrt{2}}\left( \ket{0}\ket{\psi_2} + \ket{1}\ket{\phi_2}\right) \\
= \frac{1}{2} \left( \ket{\psi_1}\ket{00}\ket{\psi_2} + \ket{\psi_1}\ket{01}\ket{\phi_2} \right. \\ \left. + \ket{\phi_1}\ket{10}\ket{\psi_2} + \ket{\phi_1}\ket{11}\ket{\phi_2}\right)
\end{split}
\end{equation*}
where $\ket{\psi_i}$ and $\ket{\phi_i}$ are $k_i$-qubit states. It is straightforward to verify that a post-measurement state of the form
\begin{equation*}
\frac{1}{\sqrt{2}}\left( \ket{\psi_1} \ket{\psi_2} \pm \ket{\phi_1} \ket{\phi_2}\right)
\end{equation*}
is obtained with probability $1/2$, where the relative phase is determined by the observed detector pattern.
If $\ket{\psi_1}, \ket{\psi_2}, \ket{\phi_1}, \ket{\phi_2}$ are all product states, and $\braket{\psi_1}{\phi_1} = \braket{\psi_2}{\phi_2} = 0$, then the post-measurement state is equivalent to a $(k_1 + k_2)$-qubit GHZ state.

The Bell state analyser is a fully \emph{loss-detecting} device (also referred to as loss-tolerant~\cite{varnava2008howgood}) because both input photons must be accounted for by the detectors in order to herald a successful outcome.
When used as a partial measurement on a subset of qubits of some input state, a fully loss-detecting measurement does not impart any lossy channel on the surviving qubits~\cite{varnava2008howgood}, regardless of how lossy the measurement device is.
The success probability of the measurement is, however, degraded by loss, as it is proportional to the probability that all input photons are detected.

\subsubsection{$n$-GHZ state analyser: type-II $n$-fusion}
\label{subsubsec:ghzsm}
\begin{figure}[htp]
    \centering
    \includegraphics[width = 0.2\textwidth]{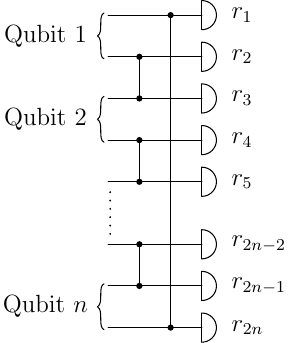}
    \caption{An $n$-GHZ analyser device composed of a linear optical network and single photon detectors, which is an $n$ qubit generalisation of Fig.~\ref{fig:BSM_circuit}'s Bell state analyser device. A measurement pattern $\bm{r}=(r_1, \dots, r_{2n})$ containing $r_i \in \{0,1\} \,\, \forall \,\,i$ indicates a destructive projection onto a specific dual rail $n$-GHZ state. Measurement patterns containing at least one $r_i=2$ correspond to projections onto known separable dual rail states.}
    \label{fig:GHZM_circuit}
\end{figure}

The design of the Bell state analyser can be extended to $n>2$ qubit inputs, as per Fig.~\ref{fig:GHZM_circuit}, to form an $n$-GHZ state analyser~\cite{pan1998greenberger, bose1998multiparticle}.

The Kraus operators for each detector pattern, when restricted to receiving $n$-qubit DR inputs, can be computed using Eq.\ \eqref{eq:evolved-projected-kraus}. 
As was the case the for 2 qubits, multiple detection patterns are associated with the same transformation.
We make use of this by grouping together Kraus operators that project onto a common state, and write them as:
\begin{equation}
\begin{split}
    K_\mathrm{S1} &= \frac{1}{\sqrt{2}} \left( \bra{0}^{\otimes n} + \bra{1}^{\otimes n} \right) \\
    K_\mathrm{S2} &= \frac{1}{\sqrt{2}} \left( \bra{0}^{\otimes n} - \bra{1}^{\otimes n} \right) \\
    K_\mathrm{Fi} & = \bra{x_i}
\end{split}
\end{equation}
where $x_i$ is the $i$th element of the set $\{ 0,1\}^n \setminus \{ 0 \dots 0, 1 \dots 1\}$, and S and F label success and failure outcomes respectively (see Appendix\ \ref{appendix:nghz-circuit-proof} for more detail).
In this form, there are $2^n$ unbiased Kraus operators, and two of them project onto an $n$-GHZ state.
The success probability for an $n$-GHZ analyser can therefore be shown to be $P_\mathrm{S} = 1/2^{n-1}$.
An experimentally friendly feature of the successful detection patterns is that they only involve $r_i\in \{0,1\}$, i.e. at most one photon present at each detector.
This means that, discounting the possibility of errors occurring from, for example, inputs containing $N>n$ photons, photon number resolving detectors are not required if only successful events are useful.
However, if information obtained from the failure events is also required, then photon number resolution of up to and including two photons is necessary.

The successful operation of an $n$-GHZ analyser can be represented by the ZX diagram
\begin{equation}\label{eq:nghzm-zx}
\vcenter{\hbox{\resizebox{0.08\textwidth}{!}{\includegraphics{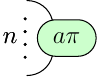}}}}
\end{equation}
where $a \in \{0,1\}$ is determined by the observed detection pattern.

$n$-GHZ analysers can perform a generalised, $n$-qubit version of type-II fusion (i.e. type-II $n$-fusion) in the GHZ basis.
Consider an extension of the example given for the Bell state analyser.
Applying an $n$-GHZ analyser device to the isolated qubits of the state
\begin{equation*}
    \frac{1}{2^{n/2}}\bigotimes_{i=1}^n \Big( \ket{\psi_i}\ket{0} + \ket{\phi_i}\ket{1}\Big) 
\end{equation*}
where $\ket{\psi_i}$ and $\ket{\phi_i}$ are, again, $k_i$-qubit states.
The post-measurement state is given by
\begin{equation*}
\frac{1}{\sqrt{2}}\left( \bigotimes_{i=1}^n \ket{\psi_i} \pm \bigotimes_{i=1}^n \ket{\phi_i} \right)
\end{equation*}
where the relative phase depends on the observed detector pattern. This is equivalent to a $(\sum_i k_i )$-qubit GHZ state if $\ket{\psi_i}, \ket{\phi_i}$ are product states and $\braket{\psi_i}{\phi_i} = 0$ for all values of $i$,.

$n$-GHZ analysers possess the same full loss-detection property as Bell state analysers.
$n$ photons must be accounted for by the detectors to herald a valid outcome, and if the device acts on part of a larger state, the uncorrelated loss-rate for the surviving qubits is the same as for the input qubits.

\subsubsection{Auxiliary assistance}
Increasing the success probability of two-qubit entangled state analysers using auxiliary assistance has been well-studied~\cite{grice2011arbitrarily, ewert2014efficient} and experimentally demonstrated~\cite{bayerbach2023bell}.
Here we begin investigating extending this concept to larger entangled state analysers.

A restricted set of modular auxiliary devices can be built from smaller, single-qubit auxiliary devices. 
A further restriction can be made, such that only single-qubit auxiliary states of the form
\begin{equation}
    \dket{\psi_\mathrm{aux}}_k = \bigotimes_{i=1}^{m_\mathrm{a}} \dket{n_i}
\end{equation}
are considered, where $k$ labels the DR qubit that the device couples to and $m_\mathrm{a}$ is the number of auxiliary modes.
Compound auxiliary devices can be built by coupling more than one qubit to single-qubit auxiliary devices.

The Kraus operators associated with a state analyser device coupled to an auxiliary device are given by
\begin{equation}
    K_{\bm{r}} = \dket{\bm{0}} \hspace{-0.1cm} \dbra{\bm{r}} \mathcal{U}^\prime P^\prime
\label{eq:aux-kraus}
\end{equation}
where $\mathcal{U}^\prime$ is the transformation enacted by composite LON (i.e. the one formed after coupling to any auxiliary modes) and $P^\prime$ is a map onto valid input states, taking into account auxiliary input states.
For example, the projector onto $n$-qubit DR inputs, each with an associated auxiliary input state $\dket{\phi_\mathrm{aux}}$, is given by $P_{\mathrm{DR}^\prime}^{\otimes n}$ where
\begin{equation*}
    P_{\mathrm{DR}^\prime} = \dket{\phi_\mathrm{aux}} \dket{10} \hspace{-0.1cm} \dbra{10} + \dket{\phi_\mathrm{aux}} \dket{01} \hspace{-0.1cm} \dbra{01}.
\end{equation*}
and the association between multimode Fock states and DR qubits can be made according to Eq.\ \eqref{eq:qubit-fock}.

\begin{figure}[htp]
    \centering
    \includegraphics[width = 0.2\textwidth]{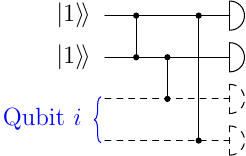}
    \caption{An auxiliary device, which we call SQA-$\beta$, can be used to increase the success probability when coupled to a pair of qubit modes (dashed) within an $n$-GHZ state analyser.}
    \label{fig:bs_aux_gadget}
\end{figure}
%

In this paper we focus on the single-qubit auxiliary device depicted in Fig.\ \ref{fig:bs_aux_gadget}.
The device can couple to any two modes of a state analyser device which have not yet interacted, but for simplicity, we choose to couple to two qubit-paired modes in this paper.
We refer to this device as SQA-$\beta$\footnote{Single Qubit Auxiliary device, which prepares an auxiliary state $\ket{\beta^-}$ (as defined in~\cite{ewert2014efficient}).}.

Coupling this to qubit 1 of the Bell state analyser (Fig.~\ref{fig:BSM_circuit}) gives an overall device that performs a modified measurement.
The Kraus operators for this device can be computed using Eq.\ \eqref{eq:aux-kraus} with
\begin{equation*}
    \begin{split}
        P^{\prime} =& \left ( \dket{1110}\hspace{-0.1cm}\dbra{10} + \dket{1101}\hspace{-0.1cm}\dbra{10} \right) \\
        & \otimes \left( \dket{10}\hspace{-0.1cm}\dbra{10} + \dket{01}\hspace{-0.1cm}\dbra{10} \right)
    \end{split}
\end{equation*}
as
\begin{equation}
\begin{split}
    K_\mathrm{S1} &= \frac{1}{\sqrt{2}} \left( \bra{00} + \bra{11} \right) \\
    K_\mathrm{S2} &= \frac{1}{\sqrt{2}} \left( \bra{00} - \bra{11} \right) \\
    K_\mathrm{S3} &= \frac{1}{2} \times \frac{1}{\sqrt{2}} \left( \bra{01} + \bra{10} \right)\\
    K_\mathrm{S4} &= \frac{1}{2} \times \frac{1}{\sqrt{2}} \left( \bra{01} - \bra{10} \right)\\
    K_\mathrm{F1} &= \frac{\sqrt{3}}{2} \bra{01} \\
    K_\mathrm{F2} &= \frac{\sqrt{3}}{2} \bra{10} 
\end{split}
\end{equation}
where we have combined similar Kraus operators (i.e. those associated with different detector patterns, but which project onto the same state) to avoid visual clutter. 
The auxiliary device has introduced outcomes associated with projection onto the Bell states $\frac{1}{\sqrt{2}} ( \ket{01} \pm \ket{10} )$ which do not exist for the unassisted Bell state analyser, without impacting the other successful outcomes.
The success probability associated with this measurement is therefore boosted relative to the unassisted version. Taking into account the weight of these extra Kraus operators gives a success probability $P_\mathrm{S} = 5/8$---an additional $1/8$ provided by the auxiliary device.

When both qubits of the Bell state analyser are each coupled to an independent copy of SQA-$\beta$, an overall Bell state analyser with $P_\mathrm{S} = 3/4$, equivalent to that introduced by Ewert and Van Loock~\cite{ewert2014efficient}, is formed.
Each copy of SQA-$\beta$ here boosts the success probability of the Bell state analyser by $1/8$.

As auxiliary assistance modifies the set of ``success" Kraus operators, it is necessary to modify the success-associated ZX diagrams acccordingly.
For the above example, the relevant ZX diagram is given by
\begin{equation}\label{eq:boosted-bsm-zx}
\vcenter{\hbox{\resizebox{0.125\textwidth}{!}{\includegraphics{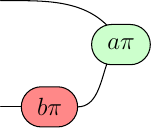}}}}
\end{equation}
where now $a,b \in \{ 0,1 \}$ each depend on the detector pattern.

The utility of applying SQA-$\beta$ carries over from the Bell state analyser to the $n$-GHZ analyser.
While it may be tempting to assume that the per-qubit boost to success probability is $1/2^{n+1}$---such that $P_\mathrm{S} = (4+n)/2^{n+1}$ if the device is applied to all $n$ qubits---the reality is subtly better.
The smallest example for which this additional boost can be witnessed is an auxiliary-assisted 4-GHZ measurement.
In this case, applying a copy of SQA-$\beta$ to every even/odd (or all) qubits introduces Kraus operators
\begin{equation}
\begin{split}
    K^\prime_\mathrm{S1} &= \frac{\sqrt{w_\mathrm{S1}}}{\sqrt{2}} \left( \bra{0101} + \bra{1010} \right) \\
    K^\prime_\mathrm{S2} &= \frac{\sqrt{w_\mathrm{S2}}}{\sqrt{2}} \left( \bra{0101} - \bra{1010} \right).
\end{split}
\end{equation}
When even/odd (all) qubits are auxiliary-coupled, $w_\mathrm{S1} = w_\mathrm{S2} = 1/256$ $(1/128)$.
Combining these contributions to success of the overall device with the others gives $P_\mathrm{S} = 25/128$ $(17/64)$.
 
We conjecture that the success probability of a $n$-GHZ analyser coupled on all qubits to a copy of SQA-$\beta$ is given by
\begin{equation}
    P_\mathrm{S} = \begin{cases}
        \frac{4+n}{2^{n+1}} + \frac{1}{2^{\frac{5n-8}{2}}} & n \text{ even}\\
        \frac{4+n}{2^{n+1}} + \frac{n}{2^{\frac{5n-9}{2}}} & n \text{ odd},
    \end{cases}
\end{equation}
although we leave formal proof of such a success probability for future work.

Modification of the ``success" Kraus operators, again, necessitates modification of the ZX diagram used to represent the device.
One can incorporate the additional outcomes by prepending X-spiders to the diagram in Eq.\ \eqref{eq:nghzm-zx} to obtain
\begin{equation}\label{eq:boosted-nghzm-zx}
\vcenter{\hbox{\resizebox{0.175\textwidth}{!}{\includegraphics{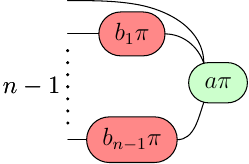}}}}
\end{equation}
where $a, b_i \in \{ 0,1 \}$ depend on the detector pattern.

The introduction of auxiliary photons and circuits changes the relationship between photon loss and success probability.
In Appendix~\ref{appendix:lossy-GHZ-analysers} we present numerically obtained data for the success probability of auxiliary-assisted 3- and 4-GHZ analyser devices under a simple loss model in which each photon sees a loss channel with loss-rate $1-\eta$.
Here we note that not all input photons need to be detected to herald a success, due to some redundancy in the associated detector patterns.

In both cases, there exist feasible loss-rate regimes in which auxiliary assistance increases the success probability.
Whether this benefit outweighs the cost of generating the auxiliary states, and the additional burden on PNRD performance, will depend on the specifics of a given platform and hardware~\cite{wein2016enhanced}.

One potential drawback of auxiliary assistance is that, unlike in the unassisted case, the successful detection patterns are not restricted to having $r_i\in \{0,1\}$.
Indeed, photon number resolution up to 5 can be required when SQA-$\beta$ is utilised.

\subsection{Type-I $n$-fusion}

\begin{figure}[htp]
    \centering
    \includegraphics[width = 0.2\textwidth]{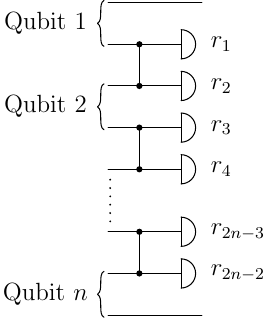}

    \caption{A linear optical heralded type-I $n$-fusion device. The map $M_n = (-1)^k \ketbra{0}{0}^{\otimes n} +  \ketbra{1}{1}^{\otimes n}$ is applied to $n$ dual-rail qubits, conditional on detecting $n-1$ photons in a pattern such that $r_i \in \{0,1\}\,\, \forall \,\, i$, and where the value of binary integer $k$ is a function of the detection pattern.}
    \label{fig:nrep-decoder}
\end{figure}

We refer to the device depicted in Fig.\ \ref{fig:nrep-decoder} as a type-I $n$-fusion device. When restricted to acting on $n$-qubit DR input states, the device implements the linear map
\begin{equation}
    M_n = (-1)^k \ketbra{0}{0}^{\otimes n} +  \ketbra{1}{1}^{\otimes n}
\end{equation}
conditional on detecting $n-1$ photons in a pattern such that $r_i \in \{0,1\}\,\, \forall \,\, i$, and where $k$ is the number of times the sub-pattern $(r_{2i} = 0, r_{2i+1}=1)$ appears in the detection pattern.

The success probability of the device, in the absence of errors, can be shown to be $P_\mathrm{S} = 1/2^{n-1}$.
This is equivalent to that of the (unassisted) $n$-GHZ analyser device, and has the benefit of leaving one qubit alive at the output rather than zero.
Full loss-detection, however, is not possessed by type-I $n$-fusion devices.
Despite this, we will later see that full loss-detection can be achieved if the device is used as part of a larger, compound measurement device.

For $n=2$, this scheme is commonly referred to simply as type-I fusion~\cite{browne2005resource-efficient}. 
The general device performs the decoder map for the $n$-qubit repetition code, up to a known local correction, and as such we can again represent the successful application of the device as a ZX diagram:
\begin{equation}\label{eq:typei-zx}
\vcenter{\hbox{\resizebox{0.08\textwidth}{!}{\includegraphics{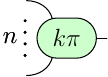}}}}
\end{equation}
Recognising this as a decoder, and therefore its inverse operation as an encoder, will help to form an intuitive link between encoded states and linear optical devices later in this paper.

While it is possible to boost the success probability of type-I fusion~\cite{bartolucci2021creation}, we are not aware of any scheme for this which uses separable, single-photon auxiliary inputs.
Because we do not consider entangled auxiliary states in this paper, we also do not study auxiliary-assisted type-I $n$-fusion devices.

\section{Flexible state generation}

In this section we combine the tools described in the previous section to highlight the flexibility and power they lend to the task of linear optical state design.

We imagine a situation in which one has a target state and would like to design a probabilistic scheme in which small entangled seed states, which can be generated using the techniques described in~\cite{bartolucci2021creation}, are combined using LONs and single photon detectors.
To achieve this, we write the state as a ZX diagram in an \emph{LO-convertible form}, which restricts the input-output relationship of the component spiders.
Allowed spiders are: (i) $0$-to-$n$, representing input seed states; (ii) $n$-to-1, representing type-I $n$-fusion devices, (iii) $n$-to-0, representing $n$-GHZ analyser devices, and; (iv) 1-to-1 spiders (including Hadamards), representing deterministic single qubit unitary devices.

It is often significantly easier to work with phase-free ZX diagrams, which only correspond to a single possible outcome for every constituent type-I $n$-fusion and GHZ analyser.
One can insert the phases associated with other measurement outcomes \emph{after} rewriting of the phase-free ZX diagrams, and manipulate them such that they emerge as single-qubit gates applied to the output state.
Different single-qubit correction operations may need to be applied for different rewritten schemes generating the same target state.
With the above in mind, we will restrict ourselves to dealing exclusively with phase-free ZX diagrams in the remainder of this paper.

\subsection{LO-convertible ZX diagrams and LO scheme extraction}

Rewriting of a ZX diagram in LO-convertible form requires the elimination of $n$-to-$m$ spiders for $n,m >1$.
This can be achieved by applying the identity
\begin{equation}\label{eq:spider-rewrite-idnetity}
\vcenter{\hbox{\resizebox{0.33\textwidth}{!}{\includegraphics{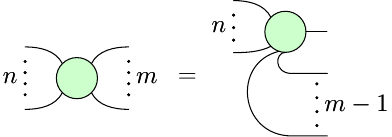}}}}
\end{equation}
wherein $m-1$ output wires become interpreted as outputs from independent Bell states, and the remaining Bell state qubits become inputs to the spider.

\begin{figure*}[htp]
    \centering
    \includegraphics[width = 0.6\textwidth]{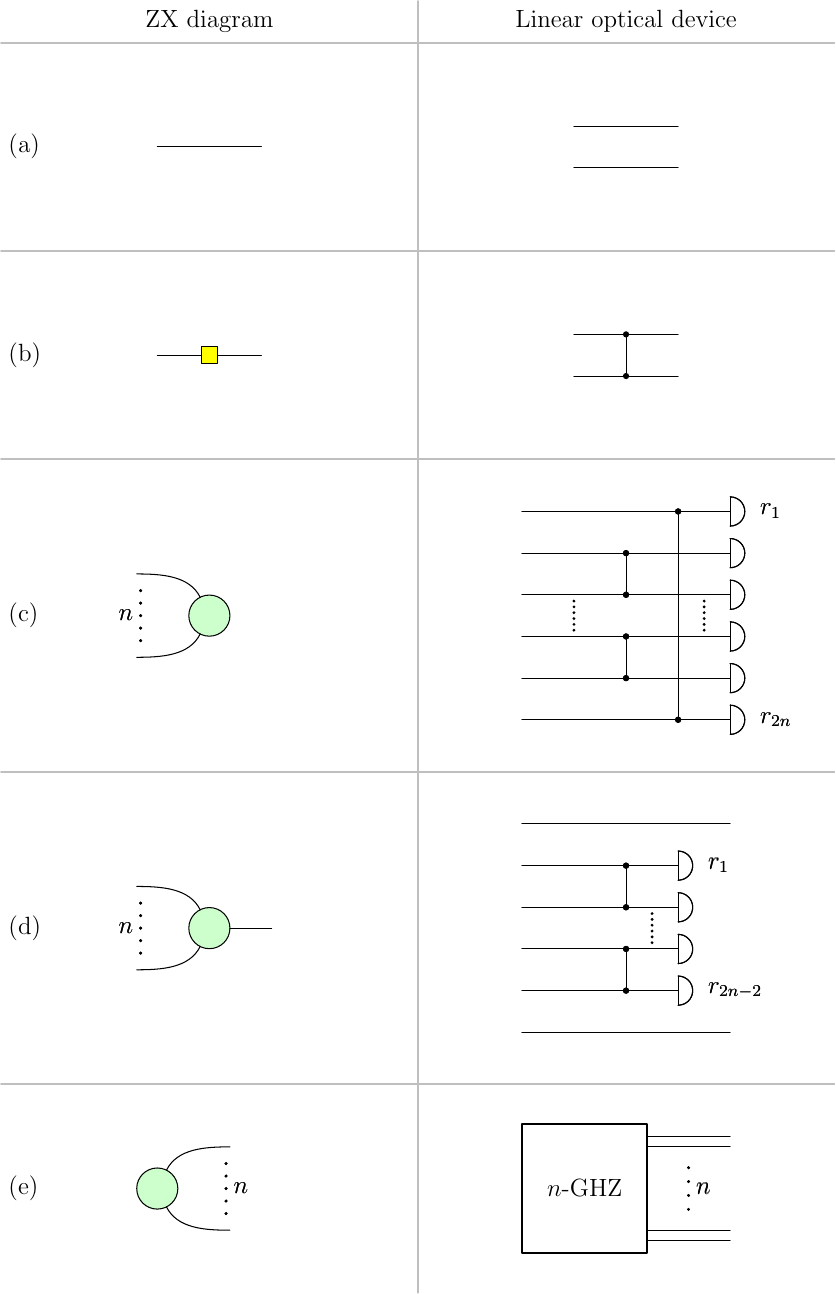}

    \caption{Informal conversion between phase-free ZX diagrams and linear optical devices. (a) Bare wires correspond to a pair of modes supporting a DR qubit. (b) Hadamards correspond to beamsplitters. (c) $n$-to-0 spiders ($n$-GHZ effects) correspond to $n$-GHZ analyser devices, conditional on obtaining a successful detector click pattern. The linear optical device can be replaced with any measurement device possessing at least one Kraus operator locally equivalent to an $n$-GHZ state, e.g. an $n$-GHZ analyser coupled to $k \leq n$ SQA-$\beta$ auxiliary devices. In the ZX diagram we omit phases, which depend on the specifics of the device and the observed detector pattern. (d) $n$-to-1 spiders correspond to type-I $n$-fusion devices, conditional on observing a valid detector pattern. The phase on the ZX diagram, which depends on the detector pattern. (e) 0-to-$n$ spiders ($n$-GHZ states) correspond to $n$-GHZ generator devices. In some cases, 0-to-2 spiders (Bell states) may be represented without the green node, i.e. as a ``cup". These devices can be of any form (e.g. from multiplexed linear optical GHZ state generator schemes, or from quantum emitters) as long as they generate the indicated state. }
    \label{fig:zx-to-lo}
\end{figure*}

A state generation scheme can be extracted from an LO-convertible ZX diagram by replacing each:
\begin{enumerate}
    \item Bare wire with two modes capable of supporting a dual rail qubit (e.g. spatially distinct waveguides or optical fibres).
    \item Hadamard with a beamsplitter.
    \item $n$-to-$0$ spider with an $n$-GHZ analyser device. If boosting is not considered, then this device is the one given in Fig.\ \ref{fig:GHZM_circuit}.
    \item $n$-to-$1$ spider with a type-I $n$-fusion device, as depicted in Fig.\ \ref{fig:nrep-decoder}.
    \item $0$-to-$n$ spider (where "cups" are equivalent to $0$-to-$2$ spiders) with a $n$-GHZ generator device.  
\end{enumerate}
These conversion rules are summarised graphically in Fig.~\ref{fig:zx-to-lo}.

There are often many possible LO-convertible ZX diagrams associated with a given target state, all of which can be transformed to one another using the graphical rewrite rules of ZX calculus.
The overall success probability of a scheme is not necessarily conserved when re-writing, especially when auxiliary assistance is considered.
Indeed, the ZX diagrams themselves contain no information about the success probability of a given scheme, but rather an indication that the scheme is valid.
Associating each spider with a LO device enables success probabilities to be calculated, and this, alongside other metrics relating to errors (loss etc.) and ease of implementation (e.g. cost of reliable large-number PNRD), allows for scheme optimisation via ZX calculus. 

A device being fully loss-detecting is often desirable, as it helps to keep the photons appearing in the final resource state from seeing too lossy a channel.
A scheme associated with an LO-convertible ZX diagram is fully loss-detecting if none of the output qubits can be traced back to the output of a type-I $n$-fusion.
In other words, resource states generated in a fully loss-detecting scheme are composed of unmeasured qubits from the seed states.
Importantly, this does not make type-I $n$-fusions incompatible with full loss-detection---if their output qubit is later measured in a GHZ analyser, the overall device is still fully loss-detecting.
As a consequence, it is almost always wasteful to create a measurement device for resource state generation exclusively from GHZ analysers.

\subsection{Encoding}

Encoding photonic qubits with error correcting codes is a way to improve performance in the presence of errors.
Often, the qubits of a resource state will be consumed in entangling measurements such as Bell state measurements as part of a computation~\cite{gimeno2015from,bartolucci2021fusion,omkar2020resource,encodedghz} or quantum communication protocol~\cite{ewert2016ultrafast,ewert2017ultrafast,lee2019fundamental}.
The success probability of these measurements can both increase~\cite{schmidt2019efficiencies} and become more tolerant to photon loss~\cite{hilaire2023linear} when the qubits are encoded with a suitable error correcting code.

Encoder maps expressed in the form of ZX diagrams have been explored for CSS~\cite{kissinger2022phase} and Clifford encoders~\cite{khesin2023graphical}, and these allow us to represent a broad class of encoded resource states as ZX diagrams.
The process of encoding a qubit within our procedure is straightforward---take the unencoded qubit and apply to it the encoder map corresponding to the desired error correcting code.
Codes can be concatenated by sequential application of encoders.

A particularly useful encoder is that for QPC$(n,m)$~\cite{bacon2006quantum}, which is a generalisation of Shor's 9-qubit code.
In one LO-convertible form, this encoder is given by
\begin{equation}\label{eq:shor_qpc}
    \begin{ZX}
        &                                                        & & & & & & \zxZ{} \ar[lllllll] \ar[end anchor = north,dddddddllllll,H={scale=.5},N-,bend right, start anchor = south west] \\
        &                                                        & & & & & & \\
        &                                                        & & & & & & \\
        &                                                        & & & & & & \\
        &                                                        & & & & & & \\
        &                                                        & & & & & & \\
        &                                                        & & & & & & \\
        &\zxZ{}  \ar[start anchor = east,urr,N-] \ar[drr,N-,start anchor = south east]   & & \zxN|{\rvdots} &[\zxNCol] \zxN|{m} & [\zxNCol]  & &\\ [\zxNRow]
        &                                                        & & & & & & \\
        &                                                        & & & & & & \\ 
        &                                                        & & & & & & \\
        & \zxN-{n \ \elevdots \ \ }                                                       & & &\zxN{}\ar[uuuuuuuuuuurrr,bend right, start anchor = north, end anchor = south,s]& & & & \\ 
        &                                                        & & & & & & \\
        &                                                        & & & & & & \\
        &                                                        & & & & & & \\
        &                                                        & & & & & & \\
        &\zxZ{}  \ar[start anchor = east,urr,N-] \ar[drr,N-,start anchor = south east]  \ar[start anchor = north,uuuuurrr,H={scale=.5},N-, end anchor = east]  & & \zxN|{\rvdots} &[\zxNCol] \zxN|{m} & [\zxNCol] & & \\ [\zxNRow]       
        &                                                        & & & & & & \\
        &                                                        & & & & & & 
    \end{ZX} 
\end{equation}

In Appendix\ \ref{appendix:encoders} we similarly show the encoder-map ZX diagrams for the 5-qubit perfect code and surface code in LO-convertible form.

\subsection{Examples}

All of the schemes obtained in this section via our procedure that involve only two-qubit operations can also be found by reasoning with type-I and type-II fusion operations on (Hadamard rotated) graph states.
While we have not devised a way to identify previously unidentifiable schemes, it is possible that our procedure makes it significantly easier and quicker to find and verify them.

While we only show the phase-free ZX diagrams for each example scheme in order to avoid clutter, we again note that the diagrams depicting the required single qubit corrections can be obtained by including them in each measurement device ZX diagram, and then propagating them to the output wires.
The state generation schemes involving seed state generators, linear optical circuits and detector arrays for the ZX diagrams in example 1 can be found in Appendix\ \ref{appendix:LO-scheme-4ghz}.
We emphasise, however, that such a scheme can be extracted from any LO-convertible diagram using the conversions presented in Fig.\ \ref{fig:zx-to-lo}.

\subsubsection{4-GHZ state}

A 4-GHZ state is perhaps the smallest resource state that one might consider generating.
It has been shown to be a valid resource state for fault tolerant, measurement based quantum computing when coupled with suitably enhanced two-qubit entangling measurements~\cite{bartolucci2021fusion}.

While there are a variety of ways to generate this state, including directly from single photons~\cite{bartolucci2021creation}, the optimal scheme will depend on the specifications of available hardware.
As such, it makes for a reasonable, simple example to highlight the ease with which different schemes can be devised.
It turns out that this example also exposes some interesting features of auxiliary-assisted entangling measurements.

We begin with the most obvious form of the ZX diagram for the 4-GHZ state
\begin{equation}
\vcenter{\hbox{\resizebox{0.025\textwidth}{!}{\includegraphics{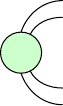}}}}
\end{equation}

Strictly speaking, this is already in an LO-convertible form, representing the direct output of a 4-GHZ state from a seed-state generator device.

The size of the seed state required can be decreased by bending the output wires via the input, as per
\begin{equation}\label{eq:zx-4ghz-bell-4ghzm}
\vcenter{\hbox{\resizebox{0.045\textwidth}{!}{\includegraphics{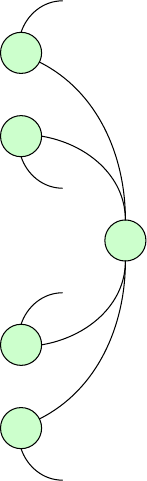}}}}
\end{equation}
corresponding to a fully loss-detecting scheme taking four Bell seed states and collectively measuring one qubit from each with a 4-GHZ analyser device, which has an unboosted success probability $P_\mathrm{S} = 1/8$.
This can be generalised to the loss-detecting generation of $n$-GHZ states---$n$ Bell states can be fused together with an $n$-GHZ measurement with base success probability $P_\mathrm{S} = 1/2^{n-1}$.

It is always possible to decompose $n$-to-0 spiders into ZX diagrams involving 2-to-1 and 2-to-0 spiders, by iteratively applying the inverse of the spider fusion rule.
Using this on \eqref{eq:zx-4ghz-bell-4ghzm} results in
\begin{equation}\label{eq:zx-4ghz-bell-bsm}
\vcenter{\hbox{\resizebox{0.07\textwidth}{!}{\includegraphics{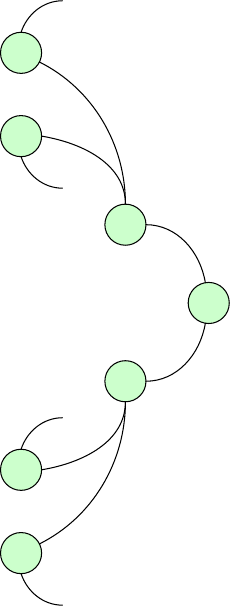}}}}
\end{equation}
where the base success probability of measurement remains unchanged at $P_\mathrm{S} = 1/8$. 
A glance at the schemes extracted from these two ZX diagrams (see Appendix~\ref{appendix:LO-scheme-4ghz}) reveals the reason for this; they are the same device.
In the absence of auxiliary assistance, all $n$-GHZ analysers can, in fact, be interpreted as trees of connected 2-qubit measurement devices.
$n$-GHZ states can, therefore, be generated in a fully loss detecting scheme using Bell seed states and standard type-I and type-II fusion devices.
Visually breaking schemes down into smaller, connected devices can be useful for identifying how groups of these sub-devices can be multiplexed independently in order to ease demands on switching networks.

Comparing the schemes in \eqref{eq:zx-4ghz-bell-4ghzm} and \eqref{eq:zx-4ghz-bell-bsm} when auxiliary assistance is considered reveals an interesting difference between previously studied boosting schemes and assisted GHZ analysers.
Consider adapting scheme \eqref{eq:zx-4ghz-bell-4ghzm} by coupling a copy of SQA-$\beta$ independently to qubits 1 and 3 of the 4-GHZ analyser, which involves 4 auxiliary photons and modes and has success probability (discounting errors and the probability of generating the seed states) of $P_\mathrm{S} = 25/128$.
This does not have an equivalent assisted scheme of the form of \eqref{eq:zx-4ghz-bell-bsm}---the auxiliary assistance would need to be spread across the two layers of the device, and the state between layers is not confined to the subspace of dual rail qubits.
A slightly different assisted scheme which is more naturally suited to \eqref{eq:zx-4ghz-bell-bsm} involves coupling SQA-$\beta$ to each qubit of the final Bell analyser, and achieves a success probability of $P_\mathrm{S} = 3/16$.
Beyond this, there is no obvious way to boost \eqref{eq:zx-4ghz-bell-bsm} further using single photon auxiliary inputs. In contrast, it is straightforward to boost \eqref{eq:zx-4ghz-bell-4ghzm} further using another 4 photons and modes, by coupling additional copies of SQA-$\beta$ to qubits~2 and~4.
This achieves a success probability of $P_\mathrm{S} = 17/64$; a 41.7\% increase over the  boosted version of \eqref{eq:zx-4ghz-bell-bsm}.

While it is unsurprising that a scheme restricted to boosting the success probability of a sub-device is sub-optimal, schemes without this restriction have until now been neglected.
The moral to take from this is similar to that expressed in~\cite{bartolucci2021creation}; forcing photons to act as qubits at intermediate stages is unnecessary, and lifting this restriction can be beneficial.

One can arrive at, perhaps, the most obvious way of generating a 4-GHZ by combining smaller entangled states by applying the spider fusion rule between the left and middle layers of \eqref{eq:zx-4ghz-bell-bsm} to give
\begin{equation}\label{eq:zx-4ghz-3ghz-bsm}
\vcenter{\hbox{\resizebox{0.045\textwidth}{!}{\includegraphics{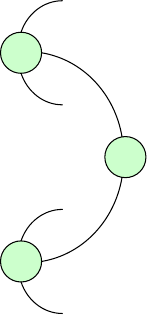}}}}
\end{equation}
Here, two 3-GHZ seed states are fused via a Bell state analyser with base success probability $P_\mathrm{S} = 1/2$ (boostable to $P_\mathrm{S} = 3/4$ when coupled to two copies of SQA-$\beta$).

\subsubsection{Six-qubit ring state}

The six-qubit ring graph state was introduced in~\cite{bartolucci2021fusion} as an improvement over the 4-GHZ state as a resource state.
A scheme has been proposed~\cite{sahay2022tailoring} for generating this state by inputting three 3-GHZ states, and performing three type-I fusion operations across them in a cyclic fashion.
This scheme can be expressed as the LO-convertible ZX diagram
\begin{equation}
\vcenter{\hbox{\resizebox{0.1\textwidth}{!}{\includegraphics{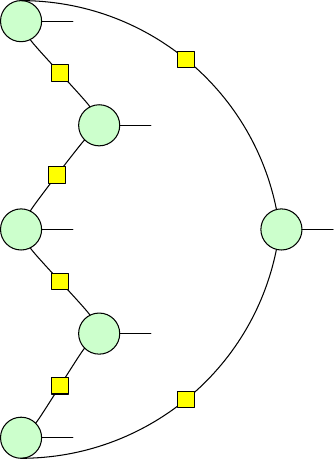}}}}
\end{equation}
which achieves a measurement success probability $P_\mathrm{S}=1/8$, but lacks full loss-detection because half of the resource state qubits are output from type-I fusion.

A simple rewriting of the above adds full loss-detection at the detriment of input resources and measurement success probability.
Bending the type-I $n$-fusion outputs via the input results in
\begin{equation}
\vcenter{\hbox{\resizebox{0.1\textwidth}{!}{\includegraphics{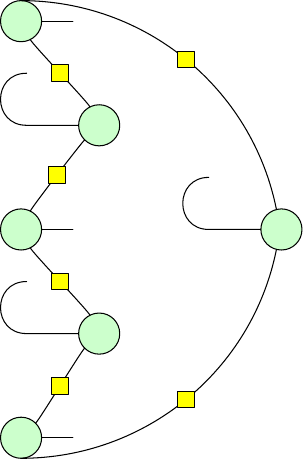}}}}
\end{equation}
which takes three 3-GHZ and three Bell seed states, and performs a $P_\mathrm{S}=1/64$ measurement via three 3-GHZ analysers.

At this point we can try to optimise the scheme to reduce the resource burden as much as possible.
Guided by the principle that $n$-GHZ analysers are less resource efficient than type-I $n$-fusions (without considering auxiliary assistance), the state can be written as
\begin{equation} \label{eq:best-6ring}
\vcenter{\hbox{\resizebox{0.09\textwidth}{!}{\includegraphics{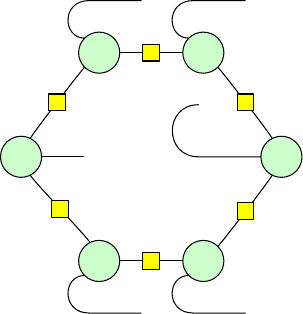}}}}
\end{equation}
While this scheme still requires a $P_\mathrm{S}=1/64$ measurement, it only needs one 3-GHZ and five Bell seed states at the input.
Furthermore, application of identity~\eqref{eq:spider-rewrite-idnetity} with $n=0$ and $m=3$ allows us to replace the 3-GHZ seed state with two Bell seed states and a type-I 2-fusion.
\begin{equation} \label{eq:best-6ring-bell}
\vcenter{\hbox{\resizebox{0.1\textwidth}{!}{\includegraphics{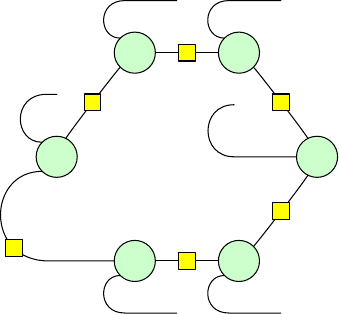}}}}
\end{equation}
This shows that the six-qubit ring state can be generated with full loss-detection from (seven) Bell seed states, with an (unboosted) $P_\mathrm{S}=1/128$ measurement.
Additional optimisation of six-ring state generation is left for future investigations.

A version of the six-qubit ring state which has enhanced error-tolerance properties has also been proposed~\cite{bartolucci2021fusion}.
In this version, each of the six qubits are encoded using the quantum parity code QPC(2,2).
By appending the QPC(2,2) encoder (i.e. Eq.~\eqref{eq:shor_qpc} with $n,m=2$) to each output wire of Eq.~\eqref{eq:best-6ring} and applying some simple rewriting, we arrive at the following scheme for generating this encoded state
\begin{equation}
\vcenter{\hbox{\resizebox{0.3\textwidth}{!}{\includegraphics{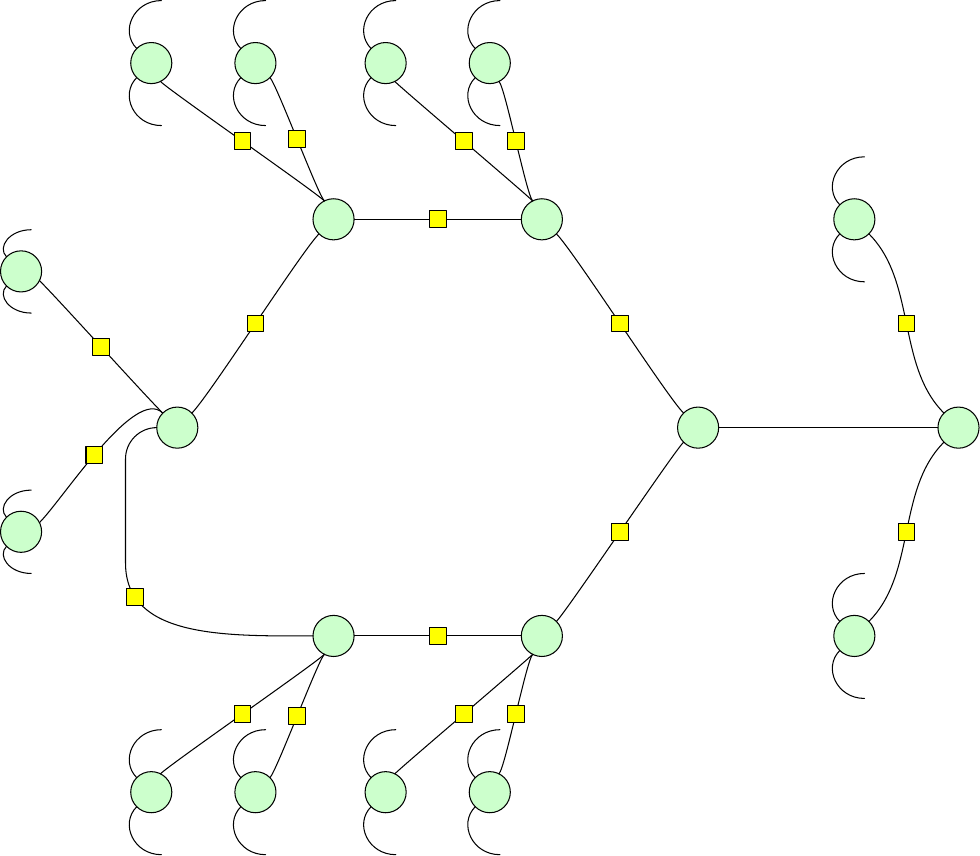}}}}
\end{equation}
Here 25 Bell seed states are combined via an overall $P_\mathrm{S}=1/2^{25}$ measurement, which can be broken down into 13 type-I 2-fusions, 5 type-I 3-fusions and a 3-GHZ analyser.

\subsubsection{Concatenated QPC(2,2) and two-qubit repetition encoded two-chain state}

This 16-qubit state is an example of a resource state presented in~\cite{encodedghz} for universal, fault tolerant measurement based quantum computation with enhanced loss-tolerance.

We start from the unencoded state---a simple two-qubit graph state.
In the form of a ZX diagram, this is given by
\begin{equation}\label{eq:2-chain}
\vcenter{\hbox{\resizebox{0.045\textwidth}{!}{\includegraphics{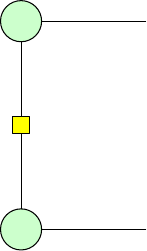}}}}
\end{equation}
The first level of encoding is a two-qubit repetition code.
Applying the encoder map for this code to Eq.\ \eqref{eq:2-chain} results in the modified ZX diagram
\begin{equation}
\label{eq:rep-2-chain}
\vcenter{\hbox{\resizebox{0.045\textwidth}{!}{\includegraphics{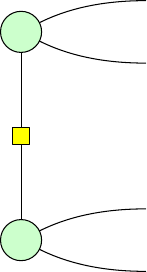}}}}
\end{equation}
To complete the encoding, QPC(2,2) is applied to each of the qubits.
As we are aiming for an overall LO-convertible ZX diagram, we choose to apply the QPC(2,2) encoder in its LO-convertible form, which is found from Eq.\ \eqref{eq:shor_qpc}.
This results in
\begin{equation}
\label{eq:orca-state-zx-1}
\vcenter{\hbox{\resizebox{0.15\textwidth}{!}{\includegraphics{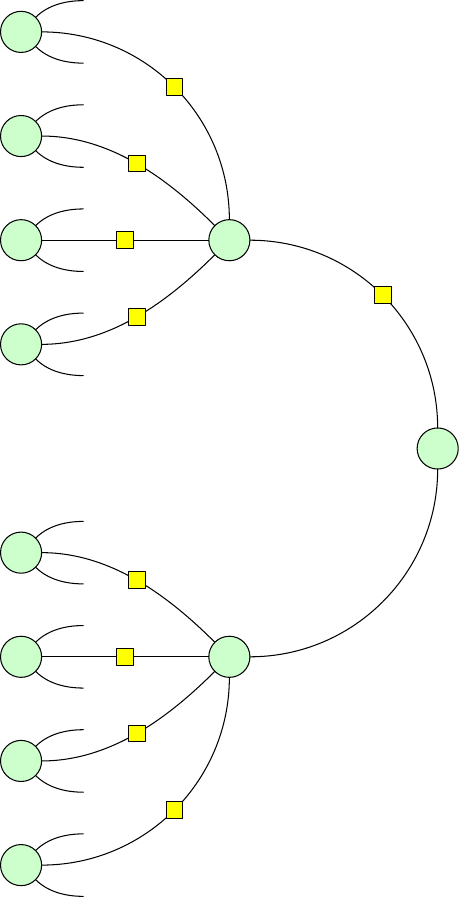}}}}
\end{equation}
which is an LO-convertible diagram for our target state.
The scheme takes 8 3-GHZ seed states, sends 2 qubits from each directly to the output and measures the rest in a 2-layer entangling measurement with basic overall success probability $P_\mathrm{S}=1/128$.
As we have seen in previous examples, we can reduce the size of seed states by application of \eqref{eq:spider-rewrite-idnetity}, such that only Bell seed states are required.
Here this results in 
\begin{equation}
\label{eq:orca-state-zx-2}
\vcenter{\hbox{\resizebox{0.15\textwidth}{!}{\includegraphics{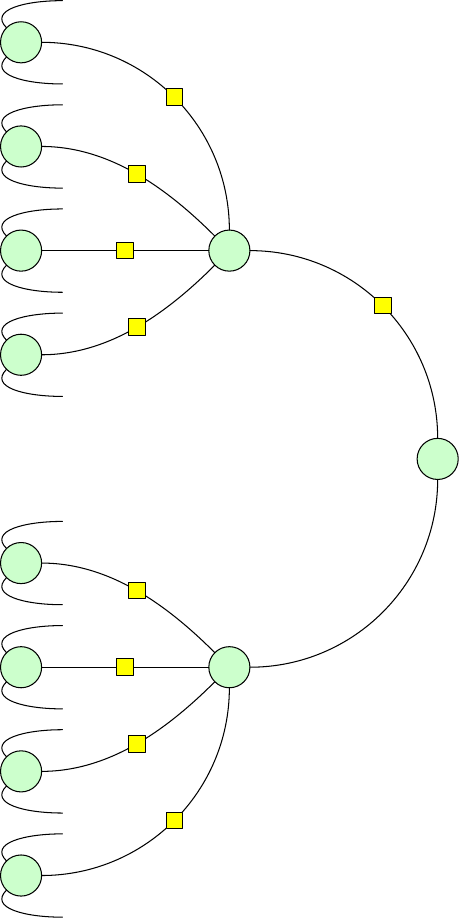}}}}
\end{equation}
which is associated with a scheme that has 16 Bell seed states, each of which provides a qubit to the output state. The remaining 16 qubits are measured in a 3-layer entangling measurement, which is a layer of type-I 2-fusions before the two layers from the scheme derived from Eq.\ \eqref{eq:orca-state-zx-1}. Overall, the measurement part of the scheme has a basic success probability $P_\mathrm{S}=1/32768$---a factor of 256 smaller than that of the previous scheme, which is due to the initial layer of type-I fusions.

In Example 1, we saw that schemes featuring larger GHZ measurements had a greater potential for boosting via SQA-$\beta$ devices.
With this in mind, we can drag the Bell effect in Eq.\ \eqref{eq:orca-state-zx-2} through to the input and change it to a Bell state, resulting in
\begin{equation}\label{eq:orca-state-zx-3}
\vcenter{\hbox{\resizebox{0.15\textwidth}{!}{\includegraphics{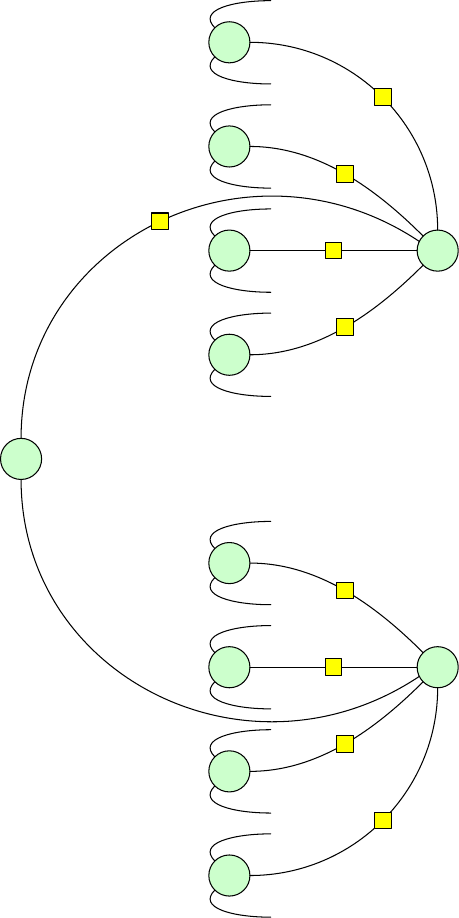}}}}
\end{equation}
This new scheme features two 5-GHZ measurements, rather than the two type-I 4-fusions and single Bell state measurement of Eq.\ \eqref{eq:orca-state-zx-2}'s scheme.
While this adaptation leads to a higher success probability measurement with full SQA-$\beta$ boosting, there is a higher resource cost to pay---16 extra auxiliary photons and an additional Bell seed state.

It might also be desirable to break the measurement from Eq.\ \eqref{eq:orca-state-zx-2} down into as many layers as possible, which may aid the design of a multiplexing scheme for the overall near-deterministic generation of the target state.
Application of the (inverse) spider fusion rule allows us to write the state generation scheme as
\begin{equation}
\label{eq:orca-state-zx-4}
\vcenter{\hbox{\resizebox{0.225\textwidth}{!}{\includegraphics{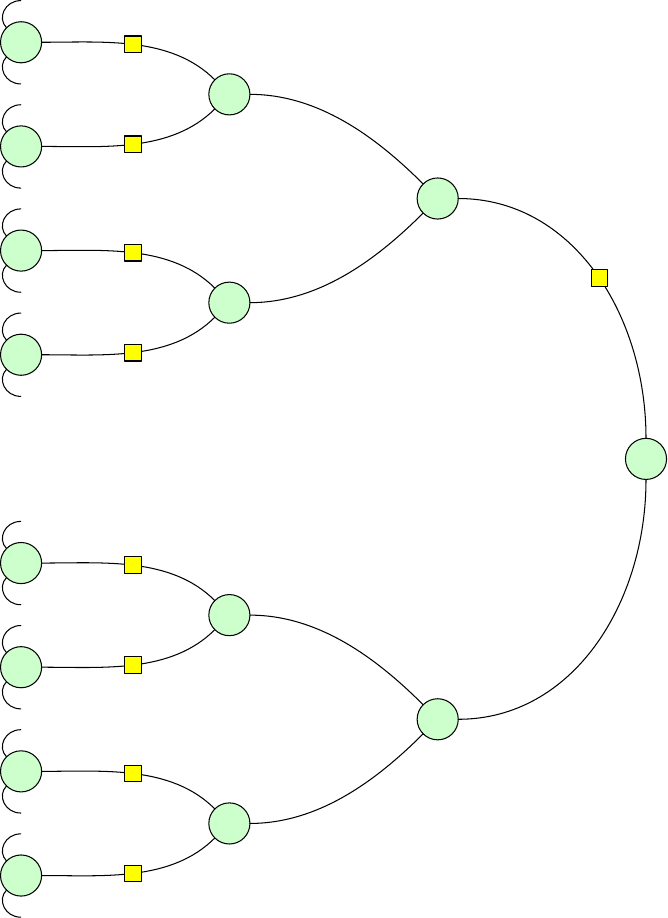}}}}
\end{equation}
which explicitly relies only on Bell seed states, type-I 2-fusions and Bell state measurements.

\section{Conclusion}

We have outlined a method which makes the task of designing schemes for preparing resource states for linear optical quantum computing easier.
The schemes output by the method are based on measuring subsets of qubits from seed states with PNRDs coupled to linear optical networks.
By assigning costs to various aspects of the schemes (e.g. measurement success probability, difficulty of preparing the seed states, realistic hardware performance etc.), the association of sub-devices with ZX diagrams allows for optimisation using ZX calculus.
As seed state generators take a generic form within this framework, schemes can be easily adapted to assess the potential benefits of utilising different seed states, e.g. 1-d graph states from quantum emitters~\cite{lindner2009proposal}.

We have provided concrete circuits for dual-rail linear optical implementation of arbitrarily sized GHZ state analysis, with and without auxiliary assistance.
While only investigating one specific, modular auxiliary state, we were able to find enhanced boosted GHZ analysers that appear to have no analog expressible in terms of boosted two-qubit measurements.
Furthermore, assessment of the performance of these boosted GHZ analysers with loss suggests that their use could be beneficial for state preparation in a loss-rate regime compatible with near-term experimental reality.
Investigating a more general form of auxiliary assistance for GHZ analysers, for example one not limited to being applied in a qubit-wise fashion, could be an interesting topic of future research.

We anticipate that adaptations and refinements can be made to our state generator design process in order to allow for further optimisation and resource savings to be made.
In particular, allowing seed and intermediate states to exist as non-qubit states, and therefore taking a more photon-centric approach (e.g.~\cite{defelice2022quantum}), could be a promising way to build upon the work presented in this paper, while retaining its ease of application and the flexibility it provides.

\section*{Acknowledgements}
The authors thank Ian Walmsley, Josh Nunn, Alex Jones, Richard Tatham and our colleagues at ORCA computing for their helpful comments. This work is partially funded by Innovate UK.


\bibliography{flexible}

\pagebreak
\onecolumngrid

\appendix

\section{Kraus operators for linear optical $n$-GHZ analysers}
\label{appendix:nghz-circuit-proof}

Here we show that linear optical $n$-GHZ analysers perform generalised measurements with Kraus operators which project onto 2 out of $2^n$ possible maximally entangled states on $n$ dual-rail encoded qubits. 

As stated in the main text, an array of $2n$ photon number resolving detectors (PNRDs) detecting $n$ photons perform a measurement with Kraus operators

\begin{equation}
    \bigotimes_{i=1}^{m}\dket{0_i}\hspace{-0.1cm}\dbra{r_i} = \dket{\bm{0}}\hspace{-0.1cm}\dbra{\bm{r}}
\end{equation}
where $\sum_i r_i = n$ and inclusion of the $i^\mathrm{th}$ mode vacuum is indicative of a destructive measurement.

When considering a measurement device comprised of a linear optical network preceding a PNRD array, the above operators are transformed according to
\begin{equation}
    \dket{\bm{0}}\hspace{-0.1cm}\dbra{\bm{r}} \mapsto \dket{\bm{0}}\hspace{-0.1cm}\dbra{\bm{r}}  \mathcal{U}(U)
\end{equation}
where $\mathcal{U}(U)$ is the transformation applied to an $n$-photon state due to the linear optical network with unitary scattering matrix $U$.

As we are dealing with dual-rail photonic qubits, we must also consider a projection to the dual rail Hilbert space from the larger multi-photon, multi-mode Hilbert space.
Starting with a projection from a single photon in two modes to the dual rail subspace,
\begin{equation}
    P_\mathrm{DR} = \ket{10}\hspace{-0.1cm}\dbra{10} + \ket{01}\hspace{-0.1cm}\dbra{01}
\end{equation}
we can write the projector onto a system of $n$ dual rail qubits as the tensor product of $n$ single qubit projectors: $P_\mathrm{DR}^{\otimes n}$.
An arbitrary $n$-qubit state in the computational dual-rail basis corresponds to a $2n$ mode Fock state
\begin{align}
    \bigotimes_{i=1}^n \ket{x_i} &= \ket{\bm{x}} \\
    &\mapsto \dket{\mathfrak{f}(\bm{x})} \\
    &= \prod_{i=1}^n \left( \hat{a}^\dagger_{2i-1}  \right)^{\delta_{x_i,0}} \left( \hat{a}^\dagger_{2i}  \right)^{\delta_{x_i,1}} \dket{\bm{0}}
\end{align}
where $x_i \in \{0,1\}$ and $\dket{\bm{0}}$ is the $2n$ mode vacuum.
The function $\mathfrak{f}(\bm{x})$ maps from length $n$ bitstrings to length $2n$ bitstrings, such that $0_i \mapsto 1_{2i},0_{2i+1}$ and $1_i \mapsto 0_{2i},1_{2i+1}$.

Using this notation, we can write
\begin{equation}
    P_\mathrm{DR}^{\otimes n} = \sum_{\bm{x} \in \{0,1\}^n} \dket{\mathfrak{f}(\bm{x})}\hspace{-0.1cm}\dbra{\mathfrak{f}(\bm{x})}.
\end{equation}

Combining the evolved $n$-photon PNRD measurement Kraus operators with a projection onto $n$ dual rail qubits gives an effective Kraus operator in the dual-rail Hilbert space given by
\begin{align}
    K^{n}_{\bm{r}} &=  \dket{\bm{0}}\hspace{-0.1cm}\dbra{\bm{r}} \, \mathcal{U} P_\mathrm{DR}^{\otimes n} \\
    &= \dket{\bm{0}}\hspace{-0.1cm}\dbra{\bm{r}} \mathcal{U} \left[ \sum_{\bm{x} \in \{0,1\}^n} \dket{\mathfrak{f}(\bm{x})}\hspace{-0.1cm}\dbra{\mathfrak{f}(\bm{x})} \right] \\
    &= \dket{\bm{0}}\hspace{-0.1cm}\dbra{\bm{r}}\hspace{-0.1cm}  \left[  \sum_{\bm{x} \in \{0,1\}^n} \mathcal{U} \dket{\mathfrak{f}(\bm{x})}\hspace{-0.1cm}\dbra{\mathfrak{f}(\bm{x})} \right].
    \label{eqn:r-projector}
\end{align}

In order to evaluate these operators, we first examine the evolution of multimode Fock states due to the linear optical network.
The linear optical circuit part of the $n$-GHZ measurement device performs the transformation
\begin{equation}
    U = U_\mathrm{BS}^{1,2n}\prod_{i=1}^{n-1}(U^{2i,2i+1}_\mathrm{BS})
\end{equation}
on the spatial modes, where $U_\mathrm{BS}^{i,j}$ is the standard beamsplitter transformation
\begin{equation}
    U_\mathrm{BS} = \frac{1}{\sqrt{2}}
    \begin{pmatrix}
    1 & 1 \\
    1 & -1 
    \end{pmatrix}
\end{equation}
between modes $i, j \in [2n]$ (i.e., $U_\mathrm{BS}^{i,j}$ is a $2n \times 2n$ matrix).
The linear optical network described by $U$ transforms each state as
\begin{align}
    \mathcal{U} \dket{\mathfrak{f}(\bm{x})}
    &= \mathcal{U} \prod_{i=1}^n \left( \hat{a}^\dagger_{2i-1}  \right)^{\delta_{x_i,0}} \left( \hat{a}^\dagger_{2i}  \right)^{\delta_{x_i,1}} \dket{\bm{0}} \\
    &= \frac{1}{2^\frac{n}{2}}\left( \hat{a}^\dagger_{1} + \hat{a}^\dagger_{2n} \right)^{\delta_{x_1, 0}}
    \left( \hat{a}^\dagger_{1} - \hat{a}^\dagger_{2n} \right)^{\delta_{x_n, 1}}
    \prod_{i=1}^{n-1} \left( \hat{a}^\dagger_{2i} + \hat{a}^\dagger_{2i+1} \right)^{\delta_{x_i, 1}}
    \left( \hat{a}^\dagger_{2i} - \hat{a}^\dagger_{2i+1} \right)^{\delta_{x_{i+1}, 0}}
    \dket{\bm{0}}
    \label{eqn:evolved-basis-state}
\end{align}
which is a sum of $2^n$ terms, where each term contains a product of $n$ creation operators.

From~\eqref{eqn:evolved-basis-state} and a given heralding pattern $\bm{r}$ we can directly deduce which input qubit states $\bm{x}$ satisfy $\dbra{\bm{r}} \mathcal{U} \dket{\mathfrak{f}(\bm{x})} \neq 0$ and therefore contribute to a particular Kraus operator.
We start by breaking up the detection pattern to highlight the pairs of modes $r_{2i}$ and $r_{2i+1}$ 
\begin{equation}
    \dbra{\bm{r}} = \dbra{r_1, r_{2n}} \bigg(\otimes_{i=1}^{n-1} \dbra{r_{2i}, r_{2i+1}} \bigg),
\end{equation}
plugging this into~\eqref{eqn:evolved-basis-state} we arrive at
\begin{align}
    \dbra{\bm{r}} \mathcal{U} \dket{\mathfrak{f}(\bm{x})} &= \frac{1}{2^\frac{n}{2}}\dbra{r_1 r_{2n}} \left( \hat{a}^\dagger_{1} + \hat{a}^\dagger_{2n} \right)^{\delta_{x_1, 0}}
    \left( \hat{a}^\dagger_{1} - \hat{a}^\dagger_{2n} \right)^{\delta_{x_n, 1}} \dket{\bm{0}} \\
    & \qquad \times 
    \prod_{i=1}^{n-1} \dbra{r_{2i},r_{2i+1}}\left( \hat{a}^\dagger_{2i} + \hat{a}^\dagger_{2i+1} \right)^{\delta_{x_i, 1}}
    \left( \hat{a}^\dagger_{2i} - \hat{a}^\dagger_{2i+1} \right)^{\delta_{x_{i+1}, 0}}
    \dket{\bm{0}}
\end{align}
where now $\dket{\bm{0}}$ is cast as a two-mode vacuum state.
We now focus our attention on the factors
\begin{equation}
    \dbra{r_{2i},r_{2i+1}} \left( \hat{a}^\dagger_{2i} + \hat{a}^\dagger_{2i+1} \right)^{\delta_{x_i, 1}}
    \left( \hat{a}^\dagger_{2i} - \hat{a}^\dagger_{2i+1} \right)^{\delta_{x_{i+1}, 0}}
    \dket{\bm{0}} = \begin{cases}
    \delta_{r_{2i},2} \delta_{r_{2i+1},0} - \delta_{r_{2i},0} \delta_{r_{2i+1},2} \quad &\text{ if } x_i = 1, x_{i+1}= 0 \\
    \delta_{r_{2i},0} \delta_{r_{2i+1},0} \quad &\text{ if } x_i = 0, x_{i+1}= 1 \\
    \delta_{r_{2i},1} \delta_{r_{2i+1},0} - \delta_{r_{2i},0} \delta_{r_{2i+1},1} \quad &\text{ if } x_i=0, x_{i+1} = 0 \\
    \delta_{r_{2i},1} \delta_{r_{2i+1},0} + \delta_{r_{2i},0} \delta_{r_{2i+1},1} \quad &\text{ if } x_i=1, x_{i+1} = 1
    \end{cases}
    \label{eq:local-patterns}
\end{equation}
with a similar expression holding for modes $1$ and $2n$.
There is a lot to unpack from this expression, however it is key to understanding the relationship between the input state and the detection patterns.
If we observe the patterns $r_{2i} = 2, r_{2i+1} = 0$ or $r_{2i} = 0, r_{2i+1} = 2$, then we know that the input state must have $x_i = 1, x_{i+1} = 0$.
Similarly, if we observe the pattern $r_{2i} = 0, r_{2i+1} = 0$, then we know that the input state satisfied $x_i = 0, x_{i+1} = 1$.
For the case where we observe the patterns $r_{2i} = 1, r_{2i+1} = 0$ or $r_{2i} = 0, r_{2i+1} = 1$, we are not able to unambiguously identify the values for $x_{i}$ and $x_{i+1}$, however we do know that they must be equal to each other.

From these observations, we can now take an observed detection pattern $\bm{r}$ and determine the unique corresponding input state $\ket{\bm{x}_{\bm{r}}}$, unless all of the observed pattern pairs are either $01$ or $10$, in which case we only know that $x_{i} = x_{i+1}$ for all $i$.
With this in mind, we proceed by considering two distinct cases for which to evaluate $P_{\bm{r}}$ :

\vspace{\baselineskip}
\noindent\textbf{Case 1:} $\bm{x} \notin \{ (0,\dots, 0), (1, \dots, 1) \}$.

The valid $\bm{r}$s associated with this case are such that not all pattern pairs $(r_{2i}, r_{2i+1})$ and  $(r_{1}, r_{2n})$ are equal to $(0,1)$ or $(1,0)$.
By~\eqref{eq:local-patterns} then, there is at least one pattern pair from $\{ (2,0), (0,2) \}$.
As $\sum r_i = n$, there must be as many $(0,0)$ pairs as pairs from $\{ (2,0), (0,2) \}$.

A consequence of the above is that there is a unique $\bm{x}$ such that 
\begin{equation}
    \dbra{\bm{r}} \mathcal{U} \dket{\mathfrak{f}(\bm{x})} \neq 0
\end{equation}
for a given valid $\bm{r}$.
This $\bm{x}$ can be inferred from a single instance of $(r_{2i}, r_{2i+1}) \in \{ (2,0), (0,2), (0,0) \}$ by the relations given in~\eqref{eq:local-patterns}.
We label this specific $\bm{x}$ as $\bm{x}_{\bm{r}}$.

Inserting $\bm{x}_{\bm{r}}$ into Eq.~\eqref{eqn:r-projector} gives
\begin{equation}
    K_{\bm{r}} = \frac{1}{2^\frac{n}{2}} \prod_i \sqrt{r_i!} \, 
    \dket{\bm{0}}\hspace{-0.1cm}\dbra{\mathfrak{f}(\bm{x}_{\bm{r}})}
\end{equation}
Furthermore, there are $2^n \prod_i \frac{1}{r_i!}$ different $\bm{r}$s for a given $\bm{x}_{\bm{r}}$ (all of which contain the same number of $r_i=2$ elements).
Combining all of these Kraus operators to form one effective Kraus operator associated with multiple outcomes gives
\begin{equation}
K_{\{ \bm{r} \}} = \dket{\bm{0}}\hspace{-0.1cm}\dbra{\mathfrak{f}(\bm{x}_{\bm{r}})}.
\end{equation}
Because $\bm{x} \in \{0,1\}^n \setminus \{(0,\dots,0), (1,\dots,1) \}$, the set of Kraus operators associated with this case, $\left\{ K_{\{ \bm{r} \}} \right\}$, amount to Kraus operators projecting onto all $n$-qubit dual-rail computational basis states except $\ket{0}^{\otimes n}$ and $\ket{1}^{\otimes n}$.

If instead we fix $\ket{\bm{x}}$, we can find the measurement patterns corresponding to a projection onto this state by associating one pattern to each term in $\mathcal{U}\dket{\mathfrak{f}(\bm{x}}$, as per Eq.\ \eqref{eqn:evolved-basis-state}.

\vspace{\baselineskip}
\noindent\textbf{Case 2:} $\bm{x} \in \{ (0,\dots, 0), (1, \dots, 1) \}$.

The cases of $00\dots0$ and $11\dots1$ can be directly computed.
They are automatically distinct from the other states because their detection patterns only consist of 0's and 1's.
By taking the linear combinations of the two states, we get the separation in the detection patterns.

\begin{equation}
\begin{split}
    \mathcal{U} \dket{\mathfrak{f}(\bm{0})}
    &= \frac{1}{2^\frac{n}{2}}\left( \hat{a}^\dagger_{1} + \hat{a}^\dagger_{2n} \right)
    \prod_{i=1}^{n-1}
    \left( \hat{a}^\dagger_{2i} - \hat{a}^\dagger_{2i+1} \right)
    \dket{\bm{0}} \\
    \\
    \mathcal{U} \dket{\mathfrak{f}(\bm{1})}
    &= \frac{1}{2^\frac{n}{2}}\left( \hat{a}^\dagger_{1} - \hat{a}^\dagger_{2n} \right)
    \prod_{i=1}^{n-1}
    \left( \hat{a}^\dagger_{2i} + \hat{a}^\dagger_{2i+1} \right)
    \dket{\bm{0}} 
\end{split}
\end{equation}

The goal is to describe precisely the relationship between the detection patterns $\bm{r}$ and the observed state $\ket{nGHZ^\pm}$. This is determined by the even- or odd-ness of the number of $01$'s that are observed in the mode pairs $(2i, 2i+1)$ plus $(1,2n)$.

\begin{equation}
\begin{split}
    \mathcal{U} \dket{\mathfrak{f}(\bm{0})}
    &= \frac{1}{2^\frac{n}{2}}\left( \hat{a}^\dagger_{1} + \hat{a}^\dagger_{2n} \right)
    \prod_{i=1}^{n-1}
    \left( \hat{a}^\dagger_{2i} - \hat{a}^\dagger_{2i+1} \right)
    \dket{\bm{0}} = 
    \frac{1}{2^\frac{n}{2}}\left( \dket{1,0}_{1,2n} + \dket{0,1}_{1,2n} \right)
    \otimes 
    \sum_{\bm{y}\in \{0,1\}^{n-1}} (-1)^{p(\bm{y})} \dket{f(\bm{y})}     \\
    \\
    \mathcal{U} \dket{\mathfrak{f}(\bm{1})}
    &= \frac{1}{2^\frac{n}{2}}\left( \hat{a}^\dagger_{1} - \hat{a}^\dagger_{2n} \right)
    \prod_{i=1}^{n-1}
    \left( \hat{a}^\dagger_{2i} + \hat{a}^\dagger_{2i+1} \right)
    \dket{\bm{0}} = 
    \frac{1}{2^\frac{n}{2}}\left( \dket{1,0}_{1,2n} - \dket{0,1}_{1,2n} \right)
    \otimes 
    \sum_{\bm{y}\in \{0,1\}^{n-1}} \dket{f(\bm{y})}
\end{split}
\end{equation}
where $p(\bm{y})$ represents the parity of the bitstring $\bm{y}$ (sum of bits modulo 2).

\begin{equation}
\begin{split}
    \frac{1}{\sqrt{2}} \mathcal{U} \big(\dket{\mathfrak{f}(\bm{0})} + \dket{\mathfrak{f}(\bm{1})} \big) &= \frac{1}{\sqrt{2^{n+1}}}
    \bigg( \dket{1,0}_{1,2n} \otimes \sum_{\bm{y}\in \{0,1\}^{n-1}} \big[ (-1)^{p(\bm{y})} + 1 \big] \otimes \dket{f(\bm{y})} \\
    & \qquad \qquad \qquad + \dket{0,1}_{1,2n} \otimes \sum_{\bm{y}\in \{0,1\}^{n-1}} \big[ (-1)^{p(\bm{y})} - 1 \big] \dket{f(\bm{y})} \bigg)\\
    &= \frac{1}{\sqrt{2^{n-1}}} \bigg(
    \sum_{\bm{y}:p(\bm{y}) = 0} \dket{1,0}_{1,2n} \dket{\mathfrak{f}(\bm{y})}
    - \sum_{\bm{y}:h(\bm{y}) = 1} \dket{0,1}_{1,2n} \otimes \dket{\mathfrak{f}(\bm{y})} \bigg) \\
    \\
    \frac{1}{\sqrt{2}} \mathcal{U} \big(\dket{\mathfrak{f}(\bm{0})} - \dket{\mathfrak{f}(\bm{1})} \big) &= \frac{1}{\sqrt{2^{n+1}}}
    \bigg( -\dket{1,0}_{1,2n} \otimes \sum_{\bm{y}\in \{0,1\}^{n-1}} \big[ (-1)^{p(\bm{y})} - 1 \big] \dket{f(\bm{y})} \\
    & \qquad \qquad \qquad + \dket{0,1}_{1,2n} \otimes \sum_{\bm{y}\in \{0,1\}^{n-1}} \big[ (-1)^{p(\bm{y})} + 1 \big] \dket{f(\bm{y})}\bigg)  \\  
    &= \frac{1}{\sqrt{2^{n-1}}} \bigg(
    \sum_{\bm{y}:p(\bm{y}) = 1} \dket{1,0}_{1,2n} \otimes \dket{\mathfrak{f}(\bm{y})}
    + \sum_{\bm{y}:p(\bm{y}) = 0} \dket{0,1}_{1,2n} \otimes \dket{\mathfrak{f}(\bm{y})} \bigg).
\end{split}
\end{equation}
The string $\bm{y}$ can be determined from an observed detection pattern $\bm{r}$ using the rule that $y_i = 0$ if $r_{2i},r_{2i+1} = 1,0$, and $y_i = 1$ if $r_{2i},r_{2i+1} = 0,1$ for $i=1,\dots,n-1$.
If the detection pattern is such that $r_1=1,r_{2n}=0$, then an even parity $\bm{y}$ indicates $\ket{nGHZ^+}$ while an odd parity $\bm{y}$ indicates $\ket{nGHZ^-}$.
On the other hand, if we have $r_1=0,r_{2n}=1$, then the relationship between the parity of $\bm{y}$ and the states $\ket{nGHZ^{\pm}}$ is reversed.

Consider, as an example, $\bm{r^\prime} = (1,1,0,1,0,\dots,1,0,0) $, which should correspond to a projection onto $\ket{nGHZ^+}$ according to the above.
We have
\begin{align}
    \dbra{\bm{r^\prime}}  \left[  \sum_{\bm{x}} \mathcal{U} \dket{\mathfrak{f}(\bm{x})}\hspace{-0.1cm}\dbra{\mathfrak{f}(\bm{x})} \right] &= \dbra{\bm{r^\prime}} \left( \mathcal{U} \dket{10\dots 10}\hspace{-0.1cm}\dbra{10\dots 10}  + \mathcal{U} \dket{01\dots 01}\hspace{-0.1cm}\dbra{01\dots 01}\right) \\
    &= 2^{-\frac{(n-1)}{2}} \left[\frac{1}{\sqrt{2}} \left( \dbra{10\dots 10} + \dbra{01\dots 01} \right)  \right]
    \label{eq:nghz-projector}
\end{align}

There are $2^{n-1}$ distinct $\bm{r}$s that give the same Kraus operator as in Eq.\ \eqref{eq:nghz-projector}.
These are associated with each of the terms in the superposition
\begin{align}
    \frac{1}{\sqrt{2}} \left( \mathcal{U} \dket{10\dots 10}  + \mathcal{U} \dket{01\dots 01} \right) = 2^{-\frac{n+1}{2}} & \left[ \left( \hat{a}^\dagger_{1} + \hat{a}^\dagger_{2n} \right) \left( \hat{a}^\dagger_{2} - \hat{a}^\dagger_{3} \right) \cdots \left( \hat{a}^\dagger_{2n-2} - \hat{a}^\dagger_{2n-1} \right) \right. \\
     &+ \left. \left( \hat{a}^\dagger_{1} - \hat{a}^\dagger_{2n} \right) \left( \hat{a}^\dagger_{2} + \hat{a}^\dagger_{3} \right) \cdots \left( \hat{a}^\dagger_{2n-2} + \hat{a}^\dagger_{2n-1} \right) \right] \dket{\bm{0}},
\end{align}
and the example of $\bm{r^\prime} = (1,1,0,1,0,\dots,1,0,0) $ given above comes from the product of the first term in each bracket.
Combining all of these Kraus operators to form one effective Kraus operator gives:
\begin{equation}
    K_{\{\bm{r} \in G_+ \}}  = \frac{1}{\sqrt{2}} \left( \dbra{10\dots 10} + \dbra{01\dots 01} \right) 
\end{equation}
where $G_+ = \left\{\bm{r}: \dbra{\bm{r}} \mathcal{U}  \dket{01}^{\otimes{n}} + \dbra{\bm{r}} \mathcal{U} \dket{10}^{\otimes{n}} \neq 0 \right\}$, i.e., the set of vectors associated with the projectors in Eq.~\eqref{eq:nghz-projector}.

Similarly, there are also $2^{n-1}$ $\bm{r}$s such that
\begin{align}
    \dbra{\bm{r}}  \left[  \sum_{\bm{x}} \mathcal{U} \dket{\mathfrak{f}(\bm{x})}\hspace{-0.1cm}\dbra{\mathfrak{f}(\bm{x})} \right] &= \dbra{\bm{r}} \left( \mathcal{U} \dket{10\dots 10}\hspace{-0.1cm}\dbra{10\dots 10}  + \mathcal{U} \dket{01\dots 01}\hspace{-0.1cm}\dbra{01\dots 01}\right) \\
    &= 2^{-\frac{(n-1)}{2}} \left[\frac{1}{\sqrt{2}} \left( \dbra{10\dots 10} - \dbra{01\dots 01} \right)  \right]
    \label{eq:nghz-projector-minus}
\end{align}
for each of them.
These $\bm{r}$s are associated with the terms in the superposition
\begin{align}
    \frac{1}{\sqrt{2}} \left( \mathcal{U} \dket{10\dots 10}  - \mathcal{U} \dket{01\dots 01} \right) = 2^{-\frac{n+1}{2}} & \left[ \left( \hat{a}^\dagger_{1} + \hat{a}^\dagger_{2n} \right) \left( \hat{a}^\dagger_{2} - \hat{a}^\dagger_{3} \right) \cdots \left( \hat{a}^\dagger_{2n-2} - \hat{a}^\dagger_{2n-1} \right) \right. \\
     &- \left. \left( \hat{a}^\dagger_{1} - \hat{a}^\dagger_{2n} \right) \left( \hat{a}^\dagger_{2} + \hat{a}^\dagger_{3} \right) \cdots \left( \hat{a}^\dagger_{2n-2} + \hat{a}^\dagger_{2n-1} \right) \right] \dket{\bm{0}}.
\end{align}
This time, the sum of the Kraus operators gives an overall effective Kraus operator projecting onto the $n$GHZ state $\frac{1}{\sqrt{2}}( \ket{0}^{\otimes n} - \ket{1}^{\otimes n} )$:
\begin{equation}
    K_{\{\bm{r} \in G_- \}}  = \frac{1}{\sqrt{2}} \left( \dbra{10\dots 10} - \dbra{01\dots 01} \right) 
\end{equation}
where $G_- = \left\{\bm{r}: \dbra{\bm{r}} \mathcal{U} \dket{01}^{\otimes{n}} - \dbra{\bm{r}} \mathcal{U} \dket{10}^{\otimes{n}} \neq 0 \right\}$.

As cases 1 and 2 cover all possible error-free outcomes, we can summarise in the following way by making associations between multimode Fock states and DR states: the ideal measurement device performs a measurement with Kraus operators
\begin{equation}
\begin{split}
    K_\mathrm{S1} &= \frac{1}{\sqrt{2}} \left( \bra{0}^{\otimes n} + \bra{1}^{\otimes n} \right) \\
    K_\mathrm{S2} &= \frac{1}{\sqrt{2}} \left( \bra{0}^{\otimes n} - \bra{1}^{\otimes n} \right) \\
    K_\mathrm{Fi} & = \bra{x_i}
\end{split}
\end{equation}
where $x_i$ is the $i$th element of the set $\{ 0,1\}^n \setminus \{ 0 \dots 0, 1 \dots 1\}$, and S and F label ``success" and ``failure", in terms of the outcome corresponding to a projection onto a maximally entangled state.


\section{Success probability performance of boosted 3-GHZ and 4-GHZ analyser devices in the presence of loss}
\label{appendix:lossy-GHZ-analysers}

\begin{figure*}[htp]
    \centering
    \subfigure[]{
    \includegraphics[width = 0.485\textwidth]{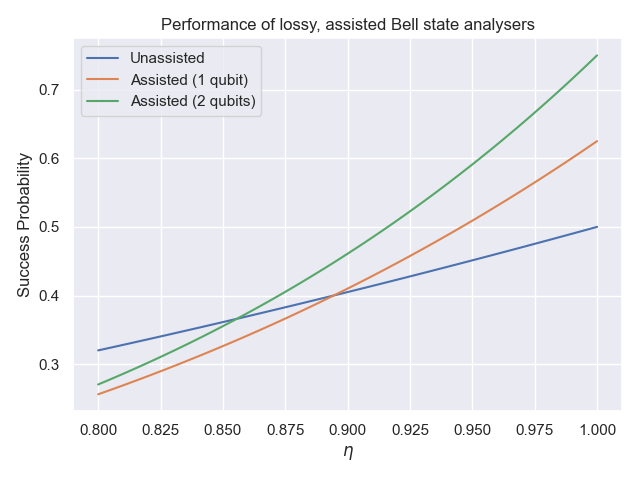}
    }\hfill
    \subfigure[]{
    \includegraphics[width = 0.485\textwidth]{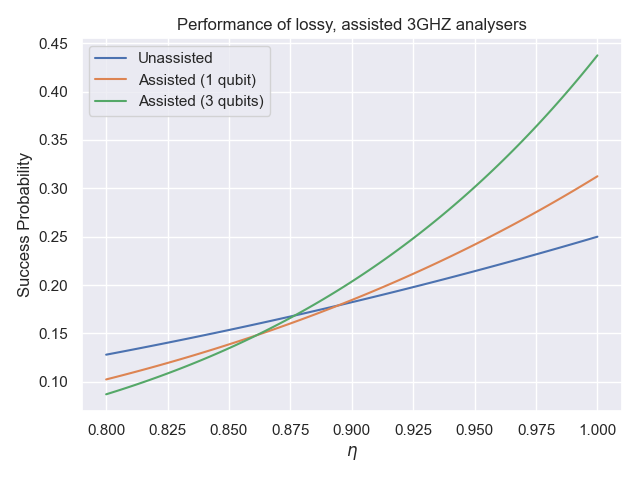}
    }
    \subfigure[]{
    \includegraphics[width = 0.485\textwidth]{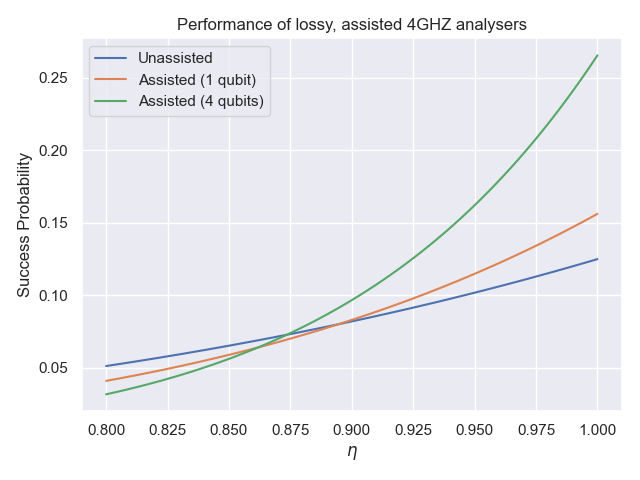}
    }
    \caption{Success probability of SQA-$\beta$ assisted (a) Bell state analysers, (b) 3-GHZ analysers, and (c) 4-GHZ analysers, for single-photon efficiency $\eta\leq1$.}
    \label{fig:aux-loss}
\end{figure*}

In a simple model in which all system and auxiliary photons see a lossy channel with loss rate $1-\eta$, we plot the success probability $P_\mathrm{S}$ for lossy, auxiliary assisted Bell state analysers, 3-GHZ analysers and 4-GHZ analysers in Fig.~\ref{fig:aux-loss}.
In each case, the assistance is given by coupling of SQA-$\beta$ to the number of qubits given on the plot.

When a copies of SQA-$\beta$ are coupled to multiple qubits of an analyser device, there exists some redundancy in the successful detector patterns that provides some loss tolerance.
In basic terms, some successful events would remain as such if one or more of the SQA-$\beta$ devices were removed.
For those events, it does not matter if those auxiliary photons get lost.
Taking this effect into account, we find the following expressions for lossy success probabilities for analysers coupled to SQA-$\beta$ devices on all qubits:
\begin{align}
 P_\mathrm{S, Bell} (\eta)&= \frac{1}{2}\eta^4 + \frac{1}{4}\eta^6 \label{eqn:lossy-aa-bsm}\\ 
 P_\mathrm{S, 3GHZ} (\eta) &= \frac{3}{8}\eta^7 + \frac{1}{16}\eta^9 \\ 
 P_\mathrm{S, 4GHZ} (\eta) &= \frac{1}{16}\eta^8 + \frac{3}{16}\eta^{10} + \frac{1}{64}\eta^{12}  
\end{align}
where\ \eqref{eqn:lossy-aa-bsm} is equivalent to the expression derived by Ewert and van Loock~\cite{ewert2014efficient}.

\section{Some encoder maps as LO-convertible ZX diagrams}
\label{appendix:encoders}

We provide examples of encoder maps explicitly in LO-convertible form to outline schemes that can realise them, and associate some notion of difficulty in applying them in a linear optics setting.

\subsection{5-qubit code}

An encoder map for the 5-qubit code can be found in~\cite{khesin2023graphical}. We restate it here as:
\begin{equation}\label{eq:5-qubit-code}
\vcenter{\hbox{\resizebox{0.33\textwidth}{!}{\includegraphics{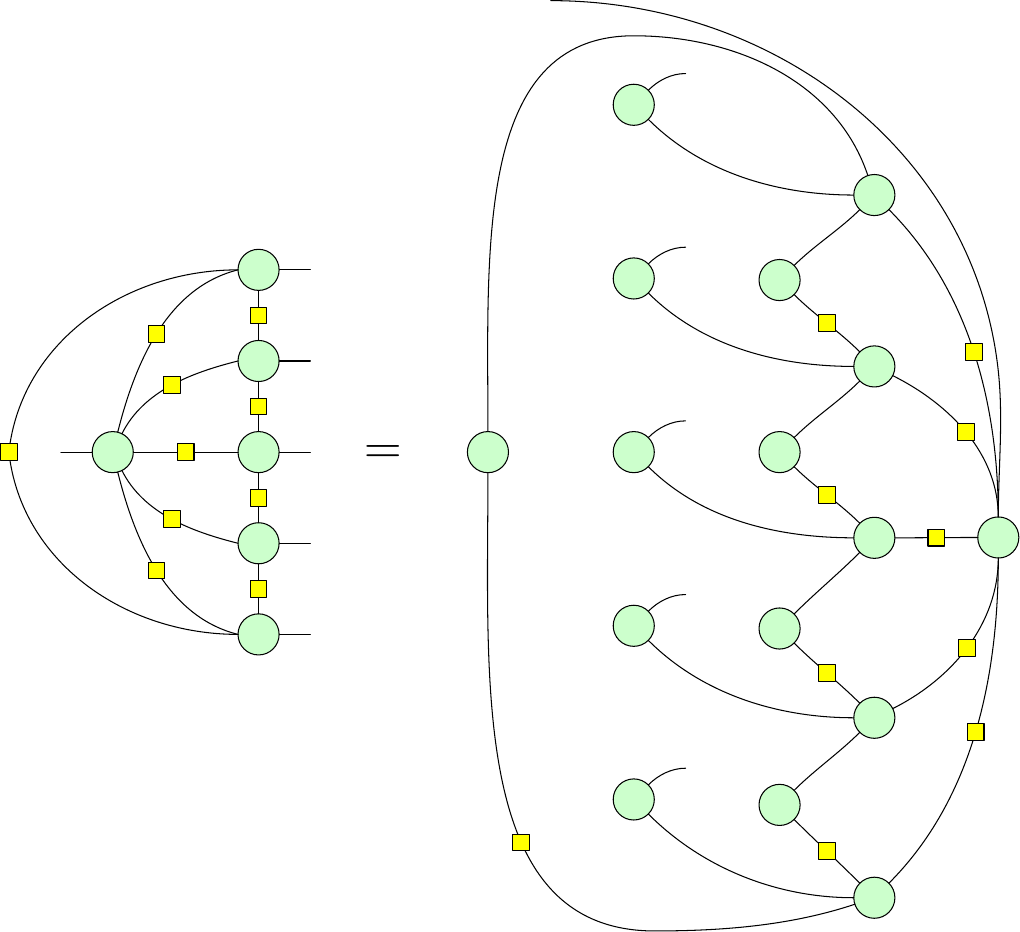}}}}
\end{equation}
where the right hand side has been rewritten in an LO-convertible form.
Due to both the cyclic and highly connected nature of the encoder, it unfortunately seems to require extensive application of the identity\ \eqref{eq:spider-rewrite-idnetity} in order to be stated in an LO-convertible form, bloating the resource requirements for implementing it.
In the form presented above, the cost of applying the encoder is 10 input Bell states, 5 type-I $3$-fusions and 1 5-GHZ analyser. Discounting the probability of generating the Bell states, the unboosted version of this device has a daunting success probability of $P_\mathrm{S} = 1/16384$. 
Furthermore, a modular scheme for multiplexing sub-devices is not immediately evident due to the sharing of Bell pairs between the type-I $3$-fusions.

\subsection{Surface code}

Any CSS encoder can be converted to two equivalent ZX diagrams using a procedure outlined in~\cite{kissinger2022phase}.
Here we pick the distance-3 surface code encoder as an example.
The form given for the encoder in~\cite{kissinger2022phase} is
\begin{equation}
\vcenter{\hbox{\resizebox{0.225\textwidth}{!}{\includegraphics{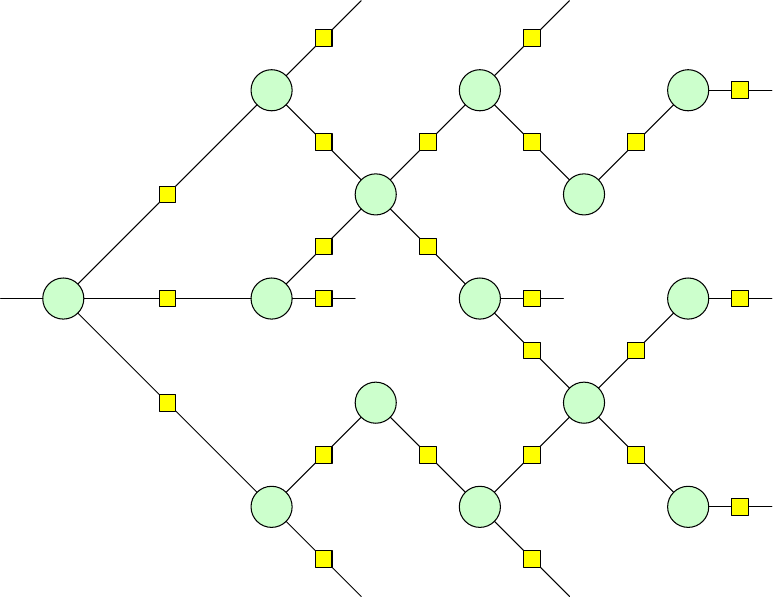}}}}
\end{equation}
which can be rewritten in an LO-convertible form
as
\begin{equation}
        \begin{ZX}
\zxNone|{}                                                                        &                                                                      & & & & & & & & &\\
\zxNone|{}                                                                        &                                                                      & & & & & & & & &\\
\zxNone|{}                                                                        &                                                                      & & & & & & & & &\\
\zxNone|{}                                                                        & \zxZ{}\ar[ddr,H={scale=.5}]\ar[ur,N-,H={scale=.5}]\ar[uur,N-,H={scale=.5}]& & & & & & & & &\\
\zxNone|{}                                                                        &                                                                      & & \zxNone|{} & & & & & & &\\
\zxNone|{} \ar[start anchor = north west,u,C] \ar[start anchor = south west,d,C]  &                                                                      &\zxZ{}\ar[start anchor = north west, end anchor = north west, ll] \ar[start anchor = south west, end anchor = south west, ll]& \zxZ{} \ar[ur,H={scale=.5}] \ar[dr,H={scale=.5}]& & & & & & & \\
                                                                                  &  \zxZ{}\ar[ur,H={scale=.5}]\ar[dr,H={scale=.5}]\ar[rrr,H={scale=.5}] & & & \zxZ{} \ar[dddl,H={scale=.5}]\ar[dddlll,H={scale=.5}] & & & & & & \zxZ{}\ar[llllulll,H={scale=.5}] \ar[dddlllllll,H={scale=.5}]\ar[uuulllllllll,N-,H={scale=.5}] \ar[dddlllllll,H={scale=.5}]\ar[uuulllulllllluul,N-]\\
                                                                                  &                                                                      & \zxNone|{} & & & & & & & &\\
                                                                                  &                                                                      & & & & & & & & &\\
                                                                                  & \zxZ{} \ar[dr,N-,H={scale=.5}] \ar[r,N-,H={scale=.5}]                                                               & & \zxZ{}\ar[r,H={scale=.5}]& \zxNone|{} & & & & & & \\
\zxNone|{}                                                                        &                                                                      & & & & & & & & &
    \end{ZX} 
\end{equation}
This version of the encoder does not make use of the identity\ \eqref{eq:spider-rewrite-idnetity} and therefore is somewhat more resource efficient to implement than that for the 5-qubit code.
In this form, it requires 2 Bell states, 4 3-GHZ states and 1 4-GHZ state as inputs, and the measurement device consists of 3 networked 4-GHZ analysers.
Discounting the generation of seed states, the success probability of the measurement device is $P_\mathrm{S} = 1/512$.


\section{LO schemes for 4-GHZ state generation}
\label{appendix:LO-scheme-4ghz}

Here we show some concrete examples of LO-based resource state generator schemes, converted from Example 1 in the main text, using the conversion rules given in Fig.\ \ref{fig:zx-to-lo}.

\begin{figure*}[htp]
    \centering
    \subfigure[]{
    \includegraphics[width = 0.3\textwidth]{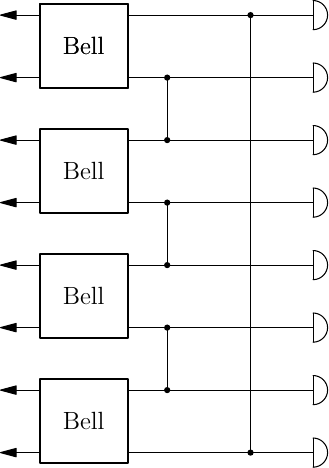}
    }
    \hspace{.75cm}
    \subfigure[]{
    \includegraphics[width = 0.34\textwidth]{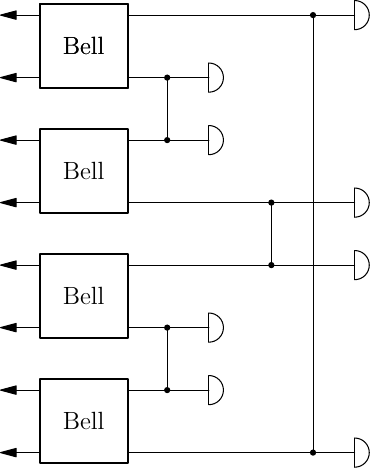}
    }
    \hspace{.75cm}
    \subfigure[]{
    \includegraphics[width = 0.3\textwidth]{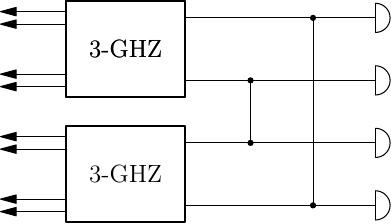}
    }
    \caption{Fully loss-detecting 4-GHZ generation schemes from seed state generators, linear optics and single photon detectors, corresponding to ZX diagrams (a) Eq.\ \eqref{eq:zx-4ghz-bell-4ghzm}, (b) Eq.\ \eqref{eq:zx-4ghz-bell-bsm} and (c) Eq.\ \eqref{eq:zx-4ghz-3ghz-bsm}. For compactness, the output state exits to the left, as indicated by the arrows. }
    \label{fig:4ghz-example}
\end{figure*}

\end{document}